\documentclass[aps,pra,twocolumn,showpacs]{revtex4-1}
\usepackage{times}
\usepackage{graphicx}
\usepackage{amsmath}
\usepackage{amsfonts}
\usepackage{amssymb}
\usepackage{color}
\usepackage{subfig}
\usepackage{soul}

\newcommand{\ket}[1]{\vert #1 \rangle} %
 
\newcommand{\ketbra}[2]{\vert #1 \rangle\langle #2 \vert} 

\begin{document}
\title{Non-Markovian continuous-time quantum walks on 
lattices with dynamical noise}
\author{Claudia Benedetti}
\email{claudia.benedetti@unimi.it}
\affiliation{Quantum Technology Lab, 
Dipartimento di Fisica, Universit\`a degli 
Studi di Milano, I-20133 Milano, Italy} %
\author{Fabrizio Buscemi}
\email{fabrizio.buscemi@unimore.it}
\affiliation{Dipartimento di Scienze Fisiche, Informatiche e Matematiche, 
Universit\`{a} di Modena e Reggio Emilia, via Campi 213/A,  Modena I-41125, Italy.}
\author{Paolo Bordone}
\email{paolo.bordone@unimore.it}
\affiliation{Dipartimento di Scienze Fisiche, Informatiche e Matematiche, 
Universit\`{a} di Modena e Reggio Emilia, via Campi 213/A,  Modena I-41125, Italy.}
\affiliation{Centro S3, CNR - Istituto Nanoscienze, Via Campi 213/A, 41125 Modena, Italy. }
\author{Matteo G. A. Paris}
\email{matteo.paris@fisica.unimi.it}
\affiliation{Quantum Technology Lab, 
Dipartimento di Fisica, Universit\`a degli 
Studi di Milano, I-20133 Milano, Italy} %
\affiliation{Istituto Nazionale di Fisica Nucleare, Sezione di Milano,
I-20133, Milano, Italy}
\affiliation{Consorzio Nazionale per la Fisica della Materia, Unit\`a
Milano Statale, I-20133 Milano, Italy}
\date{\today}
\begin{abstract}
We address the dynamics of continuous-time quantum walks on
one-dimensional disordered lattices inducing dynamical noise in the
system. Noise is described as time-dependent fluctuations of the
tunneling amplitudes between adjacent sites, and attention is focused on
non-Gaussian telegraph noise, going beyond the usual assumption of fast
Gaussian noise. We observe the emergence of two different dynamical
behaviors for the walker, corresponding to two opposite noise regimes:
{\em slow} noise (i.e. strong coupling with the environment)
confines the walker into few lattice nodes, while {\em fast} noise (weak
coupling) induces a transition between quantum and classical diffusion
over the lattice.  A phase transition between the two dynamical regimes
may be observed by tuning the ratio between the autocorrelation time of
the noise and the coupling between the walker and the external
environment generating the noise.  We also address the non-Markovianity
of the quantum map by assessing its memory effects, as well as
evaluating the information backflow to the system.  Our results suggest
that the non-Markovian character of the evolution is linked to the
dynamical behavior in the slow noise regime, and that fast noise induces
a Markovian dynamics for the walker.
\end{abstract}
\maketitle
Quantum walks (QW) are the quantum analogue of the classical random
walks \cite{kempe,venegas} and describe the propagation of a quantum
particle over a $n$-dimensional graph.  Because of their quantum nature,
which allows for quantum superposition of states and interference, QWs
show a very different behavior compared to their classical counterparts. 
This features allow one to exploit QWs for tasks that cannot be
achieved with the limited resources of classical random walks. 
Much interest has arisen around QWs especially because of
their central role in non-deterministic algorithms \cite{ambainis},
universal quantum computation \cite{childs}, transport through a graph
\cite{mulken, muelken2, darazs} and in modeling processes in biological
systems \cite{cho,nazir,odagaki}.  The generalization of random walks to
the quantum realm leads to two classes of QWs: discrete-time quantum
walks (DTQWs), where the Hilbert space of the particle position is
joined with the Hilbert space of a quantum coin \cite{aharonov}, and
continuous-time quantum walks (CTQWs), which operate only in the
position space \cite{farhi}. Both have been proved very efficient to
speedup quantum algorithms compared to their classical counterparts, 
and experimental implementation schemes have been proposed in a variety of
systems, both for DTQWs \cite{travaglione,ryan, xue, hao} and CTQWs
\cite{perets,du}.  The dynamics of a particle in discrete- and
continuous-time QW shows similar features and a formal connection  between the two classes of
quantum walks has been proved in some regimes \cite{patel,strauch}.
\par
In this paper we study  continuous-time quantum walks on a
one-dimensional graph, such as a line or a circle, i.e. quantum version
of continuous-time Markov chains \cite{farhi, tamon,blumensp}.  CTQWs on
the line are defined on a set of orthonormal position states
$\{\ket{j}\}_{k=1}^N $, where $\ket{j}$ represent a localized state of
the walker in the $j^{th}$ node of a one-dimensional lattice and $N$
is the total number of graph sites.  Due to the laws of quantum
mechanics, the quantum walker may simultaneously occupy all the lattice
nodes, with interference effects that allows the particle to propagate
faster than in the classical version.  In this paper, we will
focus on CTQW where only nearest neighbor transitions are
allowed, i.e. the particle can jump (tunnel) only to the nearest sites. 
In this scenario, the particle Hamiltonian is the
discrete Laplacian operator, i.e. the Hamiltonian describing the free
evolution
of a particle in a periodic potential:
\begin{align}
 H_0\ket{j}=2\ket{j}-\ket{j+1}-\ket{j-1}.
 \label{ctqwh}
\end{align}
The eigenvectors $\ket{\Psi_{\theta}}$ and eigenvalues $E_{\theta}$ of
the Hamiltonian \eqref{ctqwh} depend upon the choice of the boundary
conditions. In the case of periodic boundary conditions, the solutions,
found by adopting a Bloch function approach, take the expressions:
\begin{align}
 \ket{\Psi_{\theta_n}}&=\frac{1}{\sqrt{N}}
 \sum_{j=1}^N e^{-i \theta_n j}\ket{j}\label{eigenve}\\
 E_{\theta_n}&=2-2\cos\theta_n,\label{eigenva}
\end{align}
where $\theta_n=2n\pi/N$ and $n\in[1,N]$. 
Notice however that our analysis is actually independent on the boundary
conditions, since we will confine ourselves to observe the 
dynamics of the walker before it reaches the borders of the graph.  
\par
Running the quantum walks for a time $t$ means applying 
the evolution operator $U(t)=e^{-i H t}$ to an
initial state of the walker $\ket{\psi_0}$. Under the action of
Hamiltonian \eqref{ctqwh}, the quantum particle evolves with
non-classical propagation characteristics: for a localized initial state,
the QW is distributed over the lattice nodes with a highly non-Gaussian
probability distribution, showing two peaks that move away from the
initial position as time increases.  Moreover, the QWs spread more rapidly
compared to classical random walks, with a variance
$\sigma^2=\langle x^2\rangle -\langle x\rangle^2$ proportional to $t^2$
(ballistic propagation) instead of the classical diffusive propagation
$\sigma^2\propto t$.
\par
The above modeling of CTQWs is an ideal description of the diffusion of
a quantum particle over a perfect periodic potential, assuming that
neither defects nor disorder in the lattice are present.  However, in
realistic physical implementations of QWs, noise is always present
\cite{schreiber11,gius13,osellame13,kendon,sapienza}, due to fabrication
imperfections or caused by the unavoidable interaction of the walker 
with the external environment, which induces decoherence.  
Decoherence, in turn, may either suppress the propagation of the 
walker wavefunction, leading to Anderson
localization  \cite{anders58} which prevents the particle from
spreading, or it may induce a transition from quantum to classical
diffusion, thus changing the ballistic propagation of a quantum particle
to a slower diffusive spreading and destroying the interference patterns
that characterize QWs \cite{wojc}. 
\par
The effects of noise on discrete-time QWs received considerable
attention \cite{kosif,lopez,annabe,werner11, osellame14,alberti},
whereas an exhaustive analysis of the dynamics of noisy continuous-time
QWs, is still missing. As a matter of fact, the effects of static noise
or of a single impurity in the lattice, or of dynamical noise affecting
the nodes energies  have been analyzed by some authors \cite{winter07,
yin08, izaac, tamasc13}. However, full dynamical models of noise are
needed in order to give a realistic description of quantum walks,
suitable to describe the walker behavior in realistic
conditions \cite{wang14}.
\par
In this paper we address the effects of noise induced by dynamical
disorder on the behaviour of a CTQW over a one-dimensional discrete
lattice. In order to describe dynamical disorder we go
beyond the Gaussian approximation and describe noise as non-Gaussian
stochastic contribution to the tunneling amplitudes of the
Hamiltonian. In fact, quantum features in the dynamics of a walker are
mostly due to the presence of tunneling amplitudes in the interaction
Hamiltonian (as opposite to tunneling probabilities of a classical
walker) and thus adding noise to the off-diagonal elements of the
Hamiltonian allows one to assess the robustness of quantum effects to
decoherence and noise.  
\par
Besides the fundamental interest, our model is also relevant for
implementations of the QW where the imperfections arisen during the
fabrication procedure alter the coupling constant between neighboring nodes
(which could be implemented by waveguides or quantum dots), making it
not perfectly know nor constant.
We will focus on non-Gaussian noise, overcoming the widespread Gaussian
approximation for classical noise \cite{hak72}. 
Specifically, transition amplitudes
will be perturbed by random telegraph noise (RTN) which, depending on
the value of the autocorrelation time, will allow us to identify two
very different dynamical behaviors for the walker.  Moreover, we will
consider the effect of different initial conditions, specifically an
initial localized state and a Gaussian wavepacket with non-zero initial
velocity, in order to analyze both diffusion and transport phenomena on
the lattice.  In turn, analyzing the dynamical behavior of the particle
in the different working regimes is extremely relevant in the context of
reservoir engineering, as well as for noise characterization schemes
\cite{nc1,nc2,nc3,nc4,nc5}.
\par
Under the influence of noise, the walker should be 
described as an open quantum system, whose non-unitary 
evolution is influenced by an external complex
environment. In this context, addressing the memory effects of the
environment becomes a crucial issue, both from
a fundamental and an information-theoretic point of view.  We thus
complete our analysis on noisy CTQWs by connecting their
dynamical behavior with the non-Markovian (Markovian) character of the
dynamical map. {The concept of non-Markovianity of a quantum 
evolution has been discussed in terms of different analytic
properties of the corresponding dynamical map 
\cite{breuer,vasile,wolf,rhp,lun,luo,lorenzo}. The common aim of these
proposals is that to capture, possibly in a quantitative way, the 
physical mechanism making memory effects relevant for a physical
system. As such, non-Markovianity is often a useful
resource in quantum information processing \cite{chin,
vasileo,laine,huelga,bylica,manis} and our results goes in
the same direction. 
\par
As a matter of fact, different definitions and quantifiers of
non-Markovianity of a quantum map have been proposed
\cite{breuer,vasile,wolf,rhp,lun,luo,lorenzo}, many of which require an
optimization procedure which is not always feasible for systems with a
large dimensionality.  The general idea behind the concept of quantum
non-Markovianity is that the environment has memory that breaks the
time-divisibility of the dynamical map and allows information to go back
to the quantum system, e.g. recovering part of its lost coherence
\cite{lf,benedenm,chen,orieux}. The non-Markovian character of coined
QWs has been the subject of some attention \cite{peng,perez,luoma},
whereas the continuous-time case received little attention, even in the
classical case \cite{jeo13}.
\par
Here we observe that the emergence of two different (asymptotic) 
dynamical behaviors for the walker subject to non-Gaussian noise 
is linked to the presence of memory effects in the environment. 
Indeed, we show that there exists a connection between the 
autocorrelation time of the noise and the non-Markovian character 
of the dynamical map. Due to the difficulty in computing exactly 
the non-Markovianity of the evolution, we prove it 
in the presence of slow noise, whereas for fast noise we may offer
a conjecture, based on repeated numerical experiments, about the 
Markovian character of the corresponding map. \par
The paper is structured as follows: In Section \ref{sec:Model} we
introduce the model for a quantum walk on a noisy lattice, with
stochastic tunneling amplitudes. In Section \ref{sec:results} we 
present our results on the dynamics of the walker in 
the presence of noise and discuss the role of the different
noise parameters.  In Section \ref{sec:nm} we address the problem of
determining the Markovian or non-Markovian character of the dynamical
map, whereas Section \ref{sec:concl} closes the paper with some
concluding remarks.
\section{The model}\label{sec:Model}
In this Section we introduce the model for a CTQW on a one-dimensional
lattice in the presence of noise.  The physical situation we want to
describe is an implementation of the QW where, due to
imperfections arisen during the fabrication procedure of the lattice
(such as an array of waveguide) the coupling constant between
neighboring nodes is not perfectly known nor constant.  Specifically, we
describe these fabrication imperfections as stochastic time-dependent
terms in the off-diagonal elements of the Hamiltonian.  The global
Hamiltonian may thus be effectively written as:
\begin{align}
 H(t)= H_0+V(t)\label{toth}
\end{align}
where $H_0$ is the unperturbed Hamiltonian in Eq. \eqref{ctqwh}, which we
rewrite as: 
\begin{align}
H_0&=\epsilon \mathbb{I}-\sum_j \Big(\ketbra{j}{j+1}+\ketbra{j+1}{j}\Big)
 \label{ham}
\end{align}
$\epsilon$ being the on-site energy and 
$\mathbb{I}=\sum\ketbra{j}{j}$ the identity operator. The
noise contribution is described by
\begin{align}
 V(t)&=\nu\sum_j\,g_j(t)\Big(\ketbra{j}{j+1}+\ketbra{j+1}{j}\Big)
\end{align}
where the coefficients
$\{g_j(t)\}$ represent the time-dependent fluctuations of the 
tunneling amplitudes between adjacent sites of the lattices, and 
$\nu$ denotes the coupling constant between the walker and an external
environment generating the noise.  Clearly, the two matrices  do not
commute $[H_0,V(t)]\neq0$. As a consequence, they do not share a common
set of eigenvectors and Eq.s \eqref{eigenve} and \eqref{eigenva}
are not valid in the case of noisy QWs.
\par
The evolved density matrix of the particle is the ensemble average:
\begin{equation}
 \rho(t)=\langle U(t)\rho_0 U^{\dagger}(t)\rangle_{\{g_j(t)\}}
 \label{evolution}
\end{equation}
where $\langle\dots\rangle_{\{g_j(t)\}}$ denotes the average taken over all
possible realizations of the (independent) stochastic processes 
$\{g_j(t)\}$ and $U(t)$ is the
unitary evolution operator 
$$U(t)=T\,\exp\left\{-i \int_0^t\!ds\,H(s)\right\}\,,$$ where $T$ denotes 
the time-ordering operator.
\par
The noise coefficients $\{g_j(t)\}$ are stochastic classical processes
whose features describe different kinds of lattice defects and, in turn,
determine different dynamical behaviors of the walker. 
In the present work, we will focus on independent stationary processes
with autocorrelation function $$C(t)=\langle
g_j(t)g_k(0)\rangle=\delta_{jk}\chi(t)\,.$$
In particular, we focus on non-Gaussian processes and 
consider random telegraph noise, i.e. we describe the 
$\{g_j(t)\}$ as dichotomic variables which can 
switch between two values $\pm a$ with a certain switching rate $\gamma$ 
\cite{abel,zhou,bened,bened2,bened3}.
The parameter $a$ defines the strength or amplitude of the noise whereas
$\gamma$ determines its time-scale.
The probability for the fluctuator $g_j$ to switch $n$ 
times after a time $t$ follows a Poisson distribution:
\begin{equation}
 p_n(t)=\frac{(\gamma t)^n}{n!}e^{-\gamma t}.
\end{equation}
The autocorrelation function for the  process is:
\begin{align}
\chi(t)&= a^2 e^{-2\gamma t}\label{crtn}\,,
\end{align}
corresponding to a Lorentzian spectrum. 
\subsection{Non-Markovianity of the dynamical map}
The quantum map in Eq. \eqref{evolution} may give rise to  either a
Markovian or non-Markovian evolution, depending on whether  the memory
effects of the environment are negligible or they influence the walker's
dynamics. 
The non-Markovian character of the quantum evolution may be 
detected by looking at violations of equality \cite{hou} 
\begin{equation}
 T(t_2,t_0)=T(t_2,t_1)T(t_1,t_0)\label{mnm}
\,,
\end{equation}
for some triple $t_2>t_1>t_0$, where
$T$ is a universal dynamical map defined by Eq. (\ref{evolution}), i.e. 
$$T(t_b,t_a)\rho(t_a)\equiv \rho(t_b) =\langle U(t_b,t_a)\rho(t_a) 
U^{\dagger}(t_b,t_a)\rangle_{\{g_j\}}\,,$$
where $\rho(t_a)$ is and arbitrary initial state 
of the system and $t_b>t_a$. 
Whenever the dynamics may be written as a composition of two maps as in Eq. 
\eqref{mnm}, memory effects are not present and the evolution 
of the walker does not depend on its past.  
On the contrary, any violation of Eq.\eqref{mnm} provides evidence that
the future evolution depends upon all its past history, i.e. the
dynamics is non-Markovian.  In this case, in fact, given two final
states $\rho'(t_2)=T(t_2,t_0)\rho(t_0)$ and
$\rho(t_2)=T(t_2,t_1)T(t_1,t_0)\rho(t_0)$, there exists an initial state
for the system $\rho(t_0)$ such that $\rho(t_2)\neq\rho'(t_2)$.  A
nonzero distance, e.g. a nonzero trace distance, between the final 
states  $\rho(t_2)$ and $\rho'(t_2)$  can be taken as an evidence 
of non-Markovianity for quantum systems subject to classical noise.  
To support our results on the non-Markovian character of the map, 
we also look for revivals in the trace distance between a
pair of initial states subject to the same evolution, interpreted as a
signature of information backflow into the system, according to BLP
measure \cite{breuer}.
The trace distance between two quantum states 
$\rho_1$ and $\rho_2$ is defined as:
\begin{equation}
 D(\rho_1,\rho_2)=\frac{1}{2}||\rho_1-\rho_2||
 \label{tdis}
\end{equation}
where $||A||=\hbox{Tr}[\sqrt{A^{\dagger}A}]$. Whenever $D$ has 
a monotonic behavior in time, the evolution is Markovian, otherwise if 
it oscillates in time, the quantum map is non-Markovian.
\begin{widetext} 
$ $ 
\begin{figure}[h!]
\includegraphics[width=0.3\textwidth]{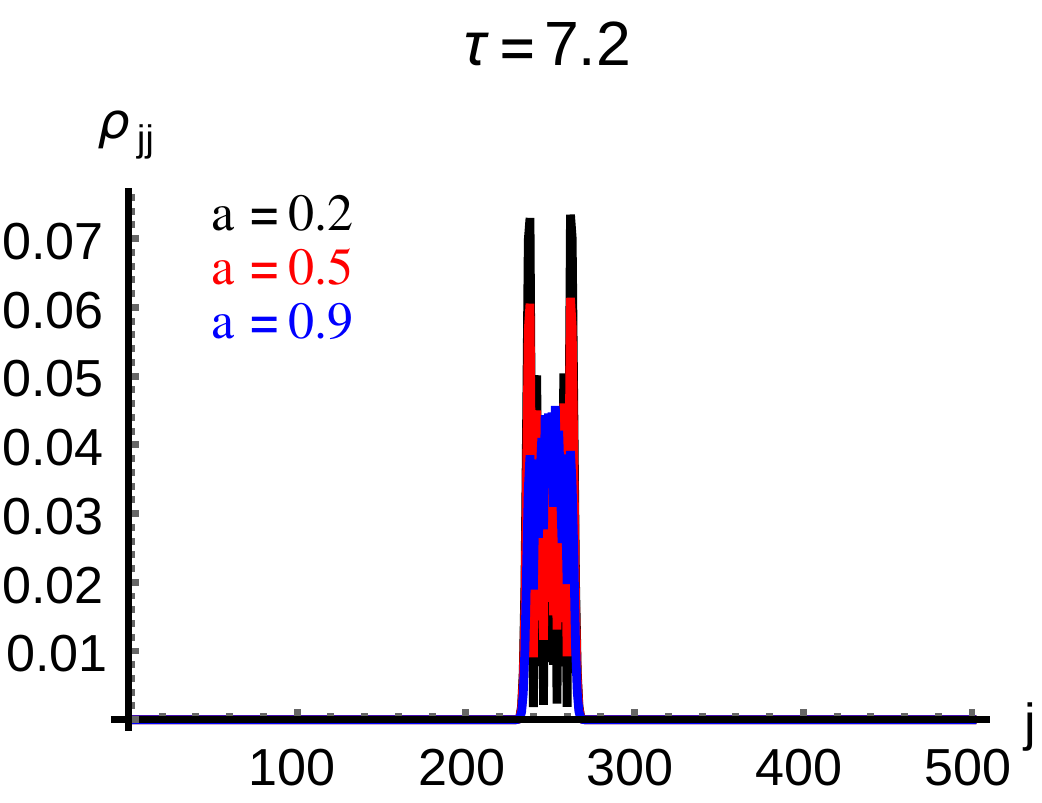}
\includegraphics[width=0.3\textwidth]{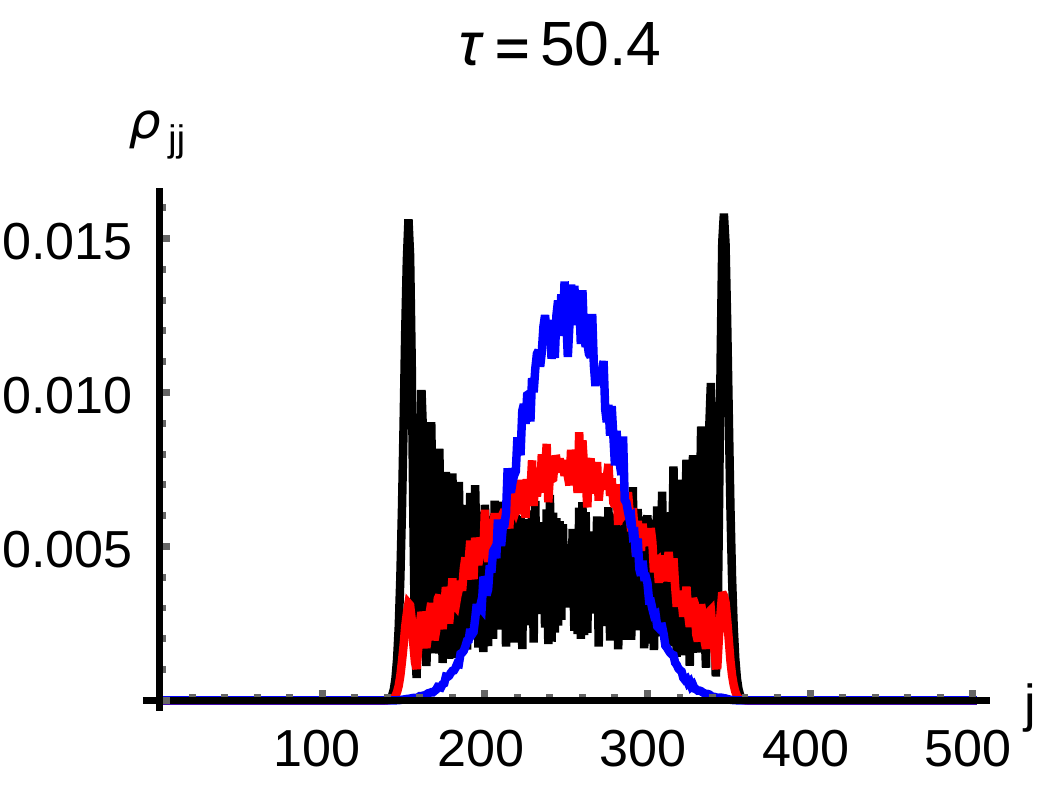}
\includegraphics[width=0.3\textwidth]{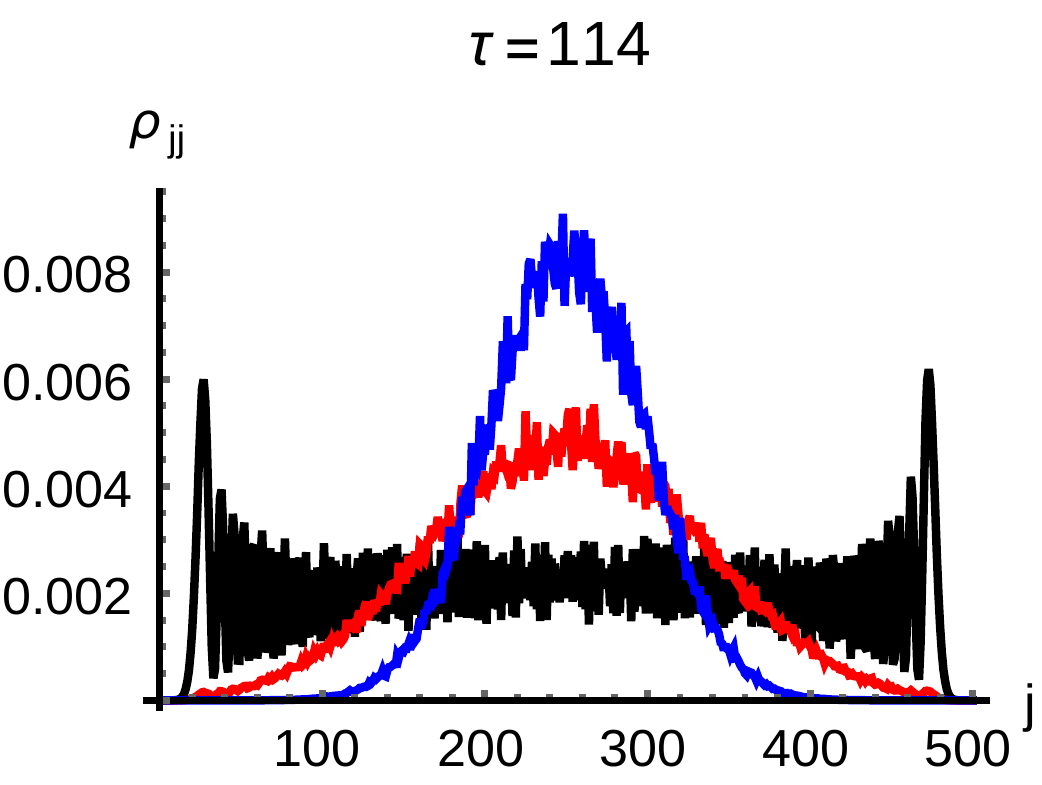}
\includegraphics[width=0.3\textwidth]{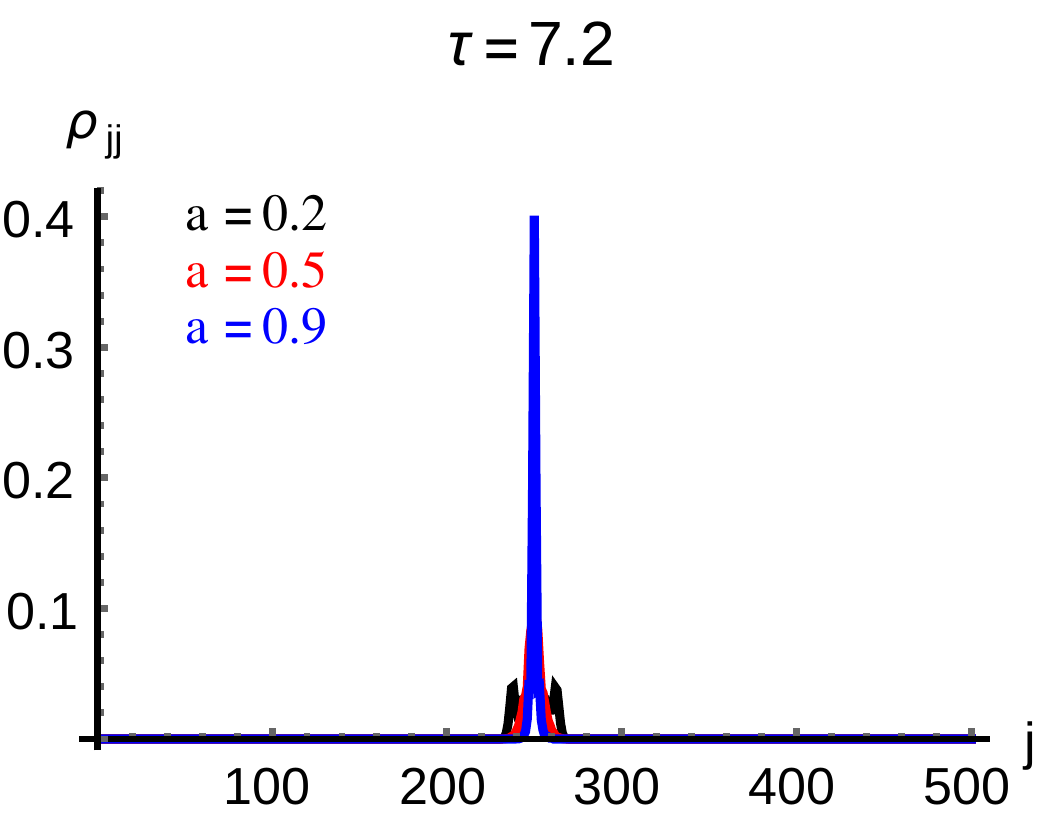}
\includegraphics[width=0.3\textwidth]{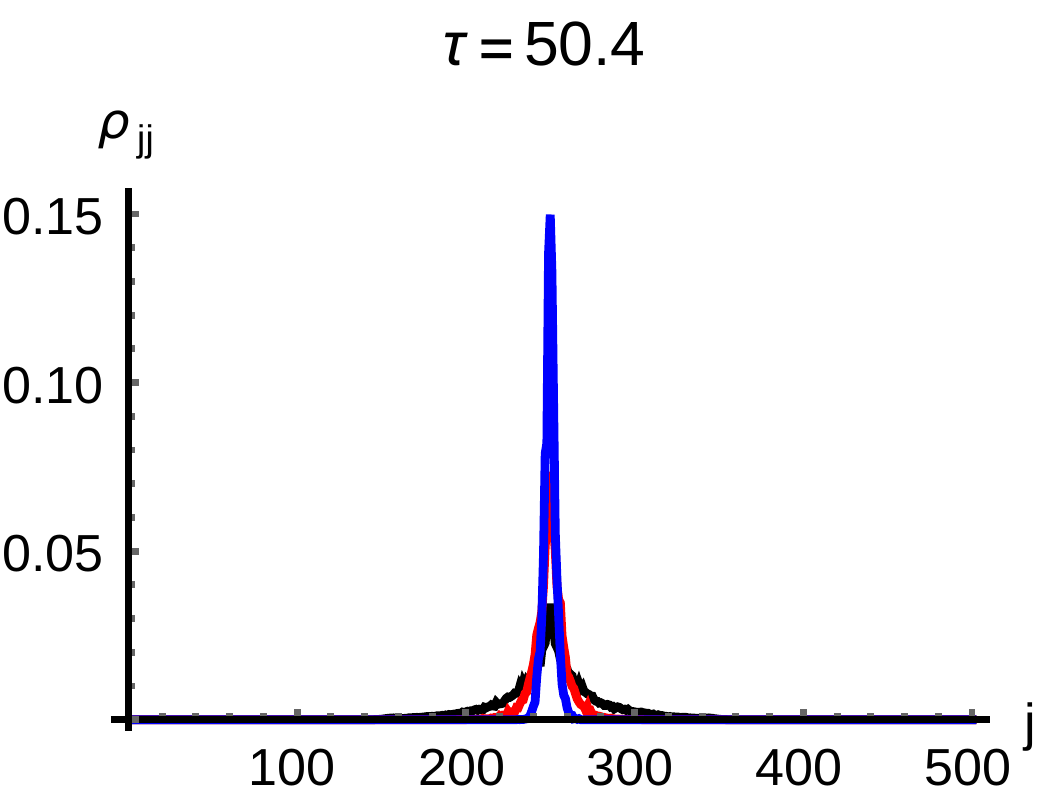}
\includegraphics[width=0.3\textwidth]{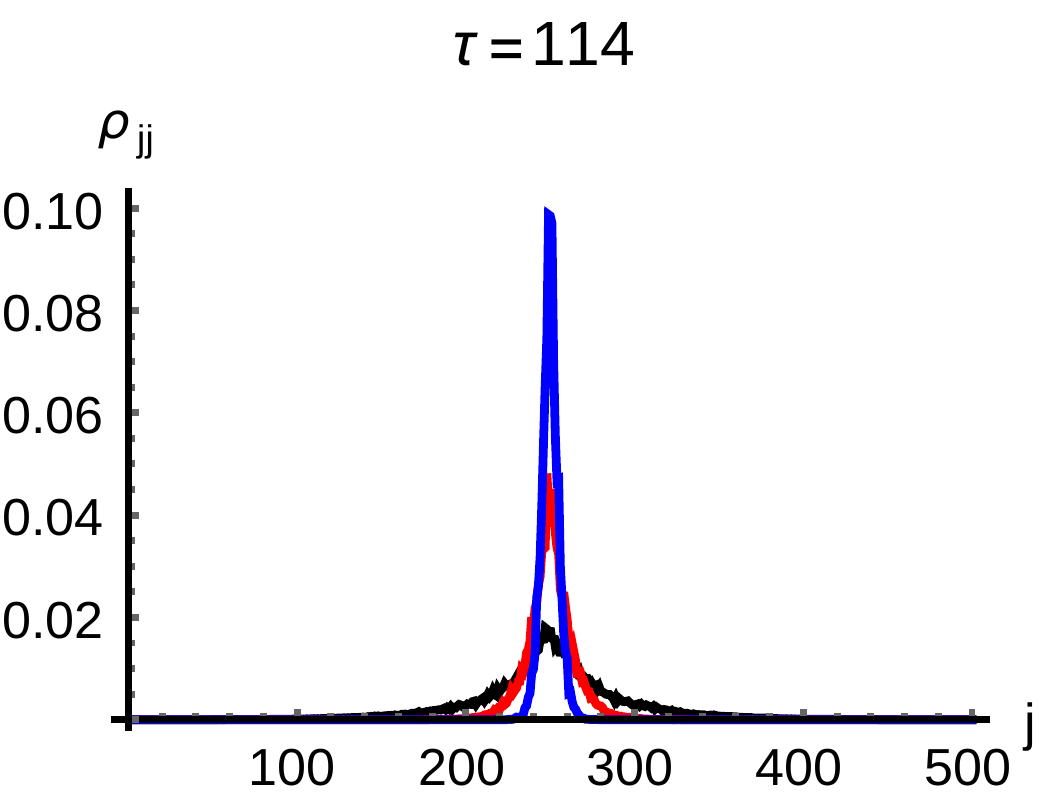}
\caption{Probability distribution of the particle over the lattice
nodes at three different  values of the interaction time
$\tau$, for a ring lattice of $N=500$ sites, subject to RTN with $\gamma=10$
(upper row) and $\gamma=0.01$ (lower row). Different values of the noise amplitude $a=0.2, 0.5$ and $0.9$ 
are represented by the black, the red (light gray) and the blue (dark gray) line respectively.
The particle is initially in
the localized state $\ket{N/2}$.} \label{rtnprof1}
\end{figure}
\end{widetext}
\par\section{Dynamics of noisy continuous-time quantum walks}
\label{sec:results}
In this section we present and discuss our results about the dynamics of
CTQW in the presence of classical noise, mimicking disorder and/or
defects in the lattice. The time evolution of the particle, described by
Eq. \eqref{evolution}, cannot be computed analytically for a large
number of nodes, so we evaluated the ensemble averages numerically after
Monte Carlo generating the values of the switching times. 
\par
In order to gain insight into the transition from quantum to classical
behavior of the walker we study the dynamics of various quantities.  
In particular, we analyze the evolution of the probability distribution
of the particle over the lattice sites using the corresponding negentropy
$N_E(t)$, the variance $\sigma^2(t)$ of the particle position and the
coherence $C(t)$ of the density matrix $\rho(t)$ as functions of time.
\begin{figure}[h!]
\includegraphics[width=0.48\columnwidth]{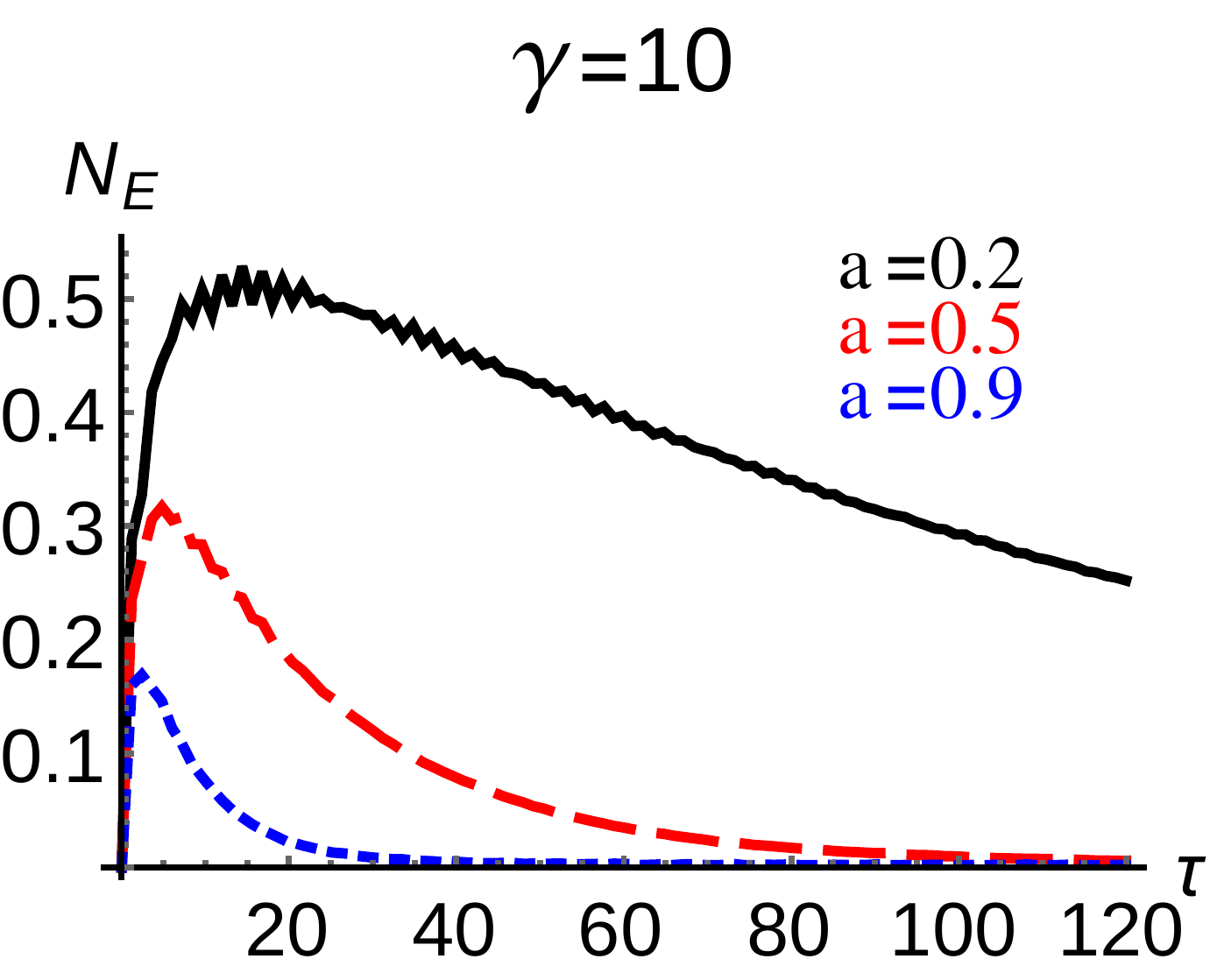}
\includegraphics[width=0.48\columnwidth]{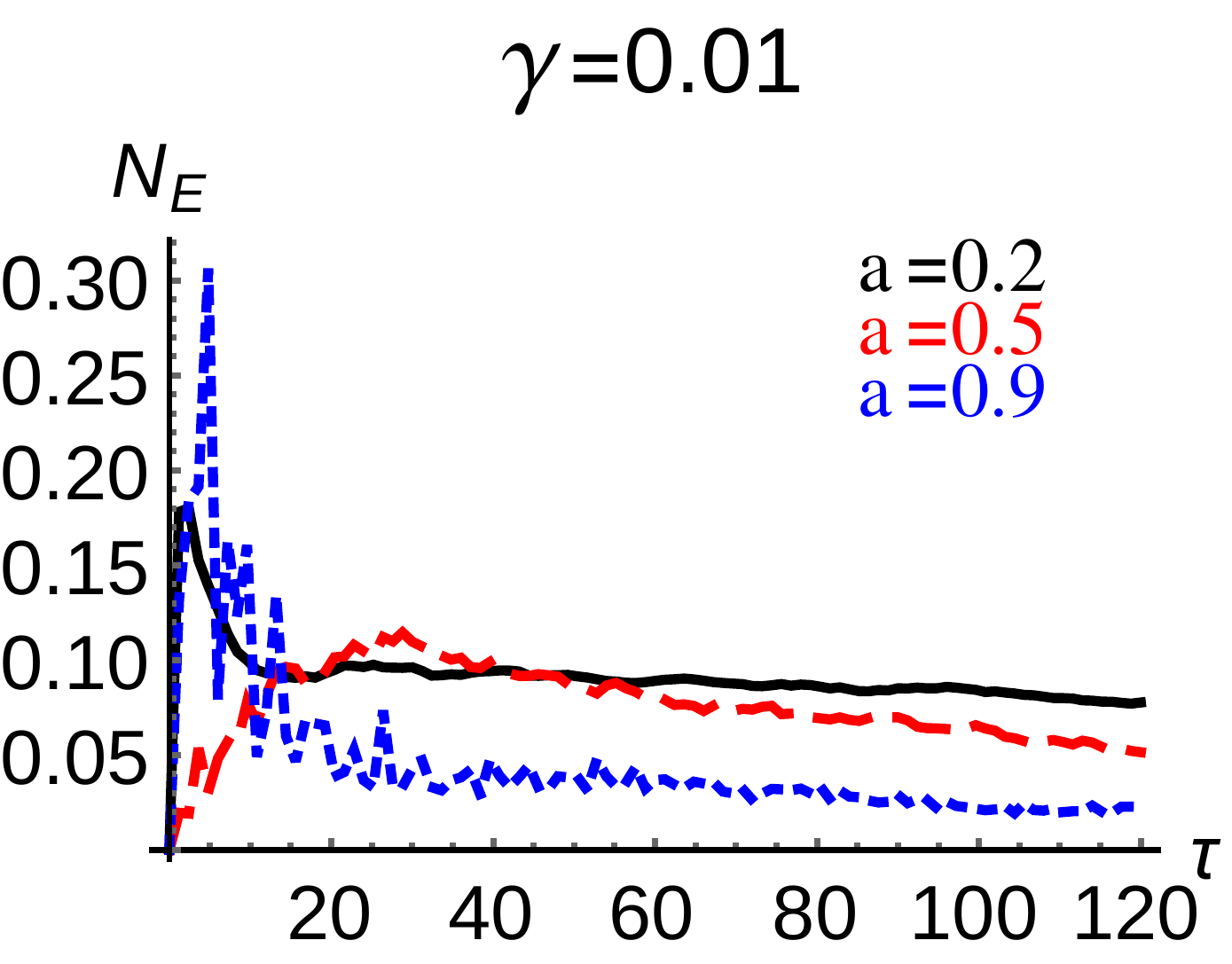}
\caption{Negentropy \eqref{negentropy} of the 
probability distribution of the particle over the lattice nodes
as a function of the interaction time $\tau$ for different values of the noise amplitude 
$a=$ $0.2$ (solid black line), $0.5$ (dashed red line)
and $0.9$ (dotted blue line), in the fast (left plot) and slow ( right plot) regime,
computed  for a QW on a 500-site lattice.} \label{negRTN}
\end{figure}
\par
The probability distribution over the lattice site corresponds to the
diagonal elements $\rho_{jj}(t)$ of the density matrix
\eqref{evolution}, while the variance is computed as $\sigma^2=\langle
x^2\rangle -\langle x\rangle^2$, where $\langle x^k\rangle = \sum_n
n^k\, \rho_{nn}$.  The negentropy $N_E$ \cite{comon} is a measure of the non-Gaussianity of
a probability distribution, i.e. it tells how much a probability
distribution deviates from a normal distribution.  The negentropy $N_E(Y)$ 
of a random variable $Y$ with distribution $p(y)$ is defined as 
difference between the Shannon entropy $H(Y_G)= - \sum_{y_G} p(y_G) 
\log p(y_G)$ of a Gaussian random variable $Y_G$ with the same 
variance of $Y$ and the Shannon entropy $H(Y)$ of $Y$
\begin{align}
N_E (Y)& = H(Y_G)-H(Y)=\nonumber\\
&=\frac{1}{2}\Big(1+\log(2\pi \sigma_y^2)\Big)+
\sum_y p(y)\,\log\big[p(y)\big]\label{negentropy}\,,
\end{align}
where $\sigma_y^2$ is the variance of both variables $Y_G$ and $Y$.
$N_E$ is always a positive quantity, unless $Y$ is Gaussian (in this 
case $N_E=0$).
\par 
Finally, the coherence of a quantum state \cite{baum} is
defined as the sum of the absolute
values of the coeherences of $\rho(t)$:
\begin{equation}
 C(t)=\sum_{k,j,k\neq j}|\rho_{kj}(t)|. 
 \label{coeren}
\end{equation}
In order to study the effect of noise on the dynamics of the 
walker, different initial conditions have been considered, including
both the case of a state $|\psi_0\rangle = |j_0\rangle$ initially
localized on a single node of the lattice, and of a Gaussian 
wavepacket with a certain width  $\Delta$ and an
initial non-zero velocity.  The tunable parameters which we can change
in order to obtain different dynamical evolutions for the walker are the
amplitudes of the noise terms $g_j(t)$, which at every instant of time 
take values $\pm a$, and the switching rate $\gamma$.
\par
In order to simplify the analysis of the dynamics, we exploit a scaling 
property of the system (with respect to the coupling $\nu$) and 
introduce the dimensionless time and the dimensionless 
switching rate as 
\begin{equation}
t \rightarrow \nu t \equiv \tau \quad \gamma\rightarrow\gamma/\nu\,.
\label{rescalp}
\end{equation}
\subsection{Localized initial state}
Let us first focus to the case where the particle is initially localized
on the central lattice site $|\psi_0\rangle=|N/2\rangle$.  Fig.
\ref{rtnprof1} shows the probability distribution of the particle over
the lattice sites at three different interaction times $\tau$ for
selected noise amplitudes $a$ in the case fast noise $\gamma=10$ (upper
row) and
slow noise $\gamma=0.01$ (lower row).
The two chosen values for the switching rates are good 
representatives of the two regimes of the RTN with 
fast ($\gamma\gg1$) and slow ($\gamma\ll1$) decaying 
autocorrelation function. In particular, 
a large value of $\gamma$ corresponds to a situation where the 
bistable fluctuators flip almost at every time step (remember that the
average number of switches in a time interval $dt$ is $n=\gamma\, dt$),
while RTN for very small values of the switching rate can be considered
an example of quasi-static (but still random) noise. 
\par 
The first fact emerging from Fig.
\ref{rtnprof1} is  that the two different regimes give rise to very
different behaviors. Under the action of fast RTN, the walker spreads
over the lattice with a probability distribution dependent on the noise
strength $a$.  The higher is the noise amplitude $a$, the stronger
is the impact of defects on the dynamics of the walker.  A transition
from quantum to classical is induced over time. The probability
distribution of the unperturbed walker, with the two peaks moving
away from the starting node, is lost as the value of $a$ is
increased: while for small values of the noise amplitude the typical
quantum behavior is still present during the time evolution, for larger
values of $a$ the interference pattern is completely lost  already at
small interaction times, and a Gaussian-like distribution centered
around the initial position arises for large times $\tau$.  The two tail
peaks are suppressed while the central part of the distribution grows,
as the value of $a$ is increased.
The situation is very different in the case of a slow noise: the
doubly-peaked distribution vanishes with increasing noise amplitudes,
but the probability distribution remains localized around
its initial position. This effect is the so-called
Anderson localization \cite{anders58}, already found for static noise
affecting the diagonal terms of the Hamiltonian \cite{keating,li, ghosh}. 
\begin{figure}[h]
\centering
\includegraphics[width=0.49\columnwidth]{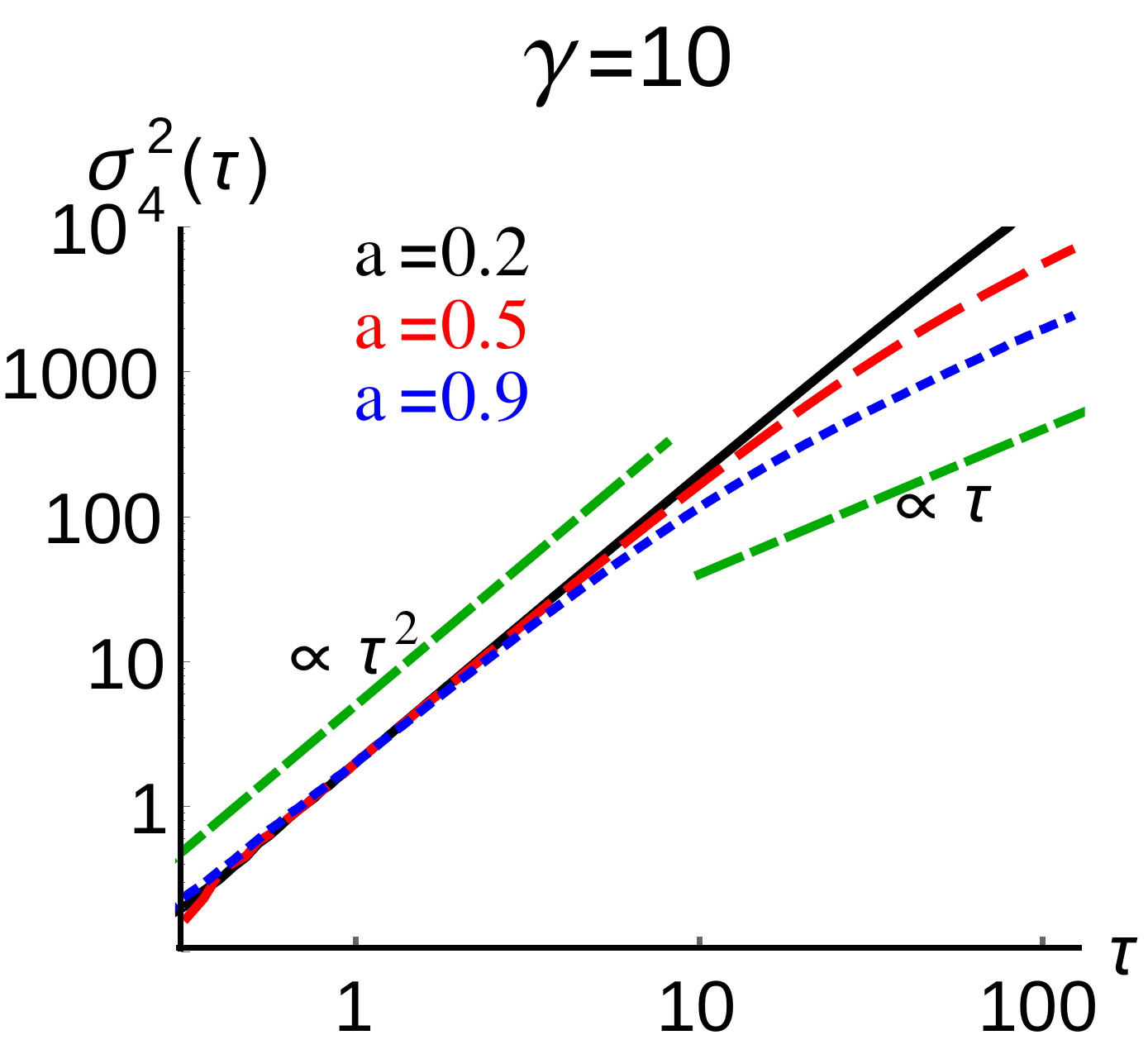}
\includegraphics[width=0.49\columnwidth]{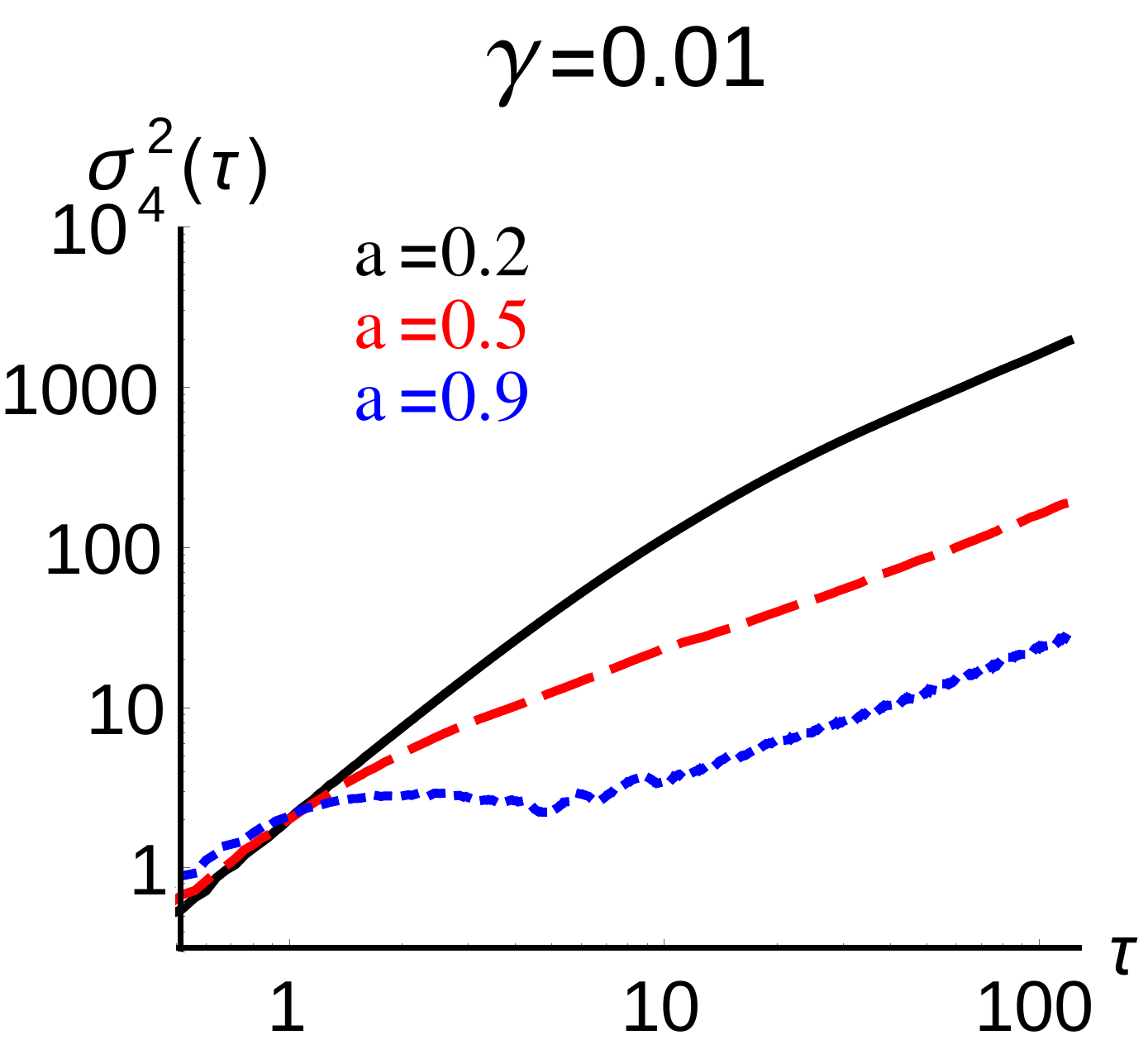}
\caption{Variance $\sigma^2(\tau)=\langle x^2 \rangle - \langle x
\rangle ^2$ of the particle position as a function of the interaction time $\tau$, for a  lattice of $N=500$ sites,
subject to RTN with $\gamma=10$ (left plot) and $\gamma=0.01$ (right
plot) for different values of the noise
amplitude $a=$ $0.2$ (solid black line), $0.5$ (dashed red line)
and $0.9$ (dotted blue line). As a guide for the eyes, the green dot-dashed curves indicate the different slopes in the 
left plot. 
The particle is initially in the localized state $\ket{N/2}$.}
\label{sigmaRTN}
\end{figure}
\begin{figure}[h]
\includegraphics[width=0.49\columnwidth]{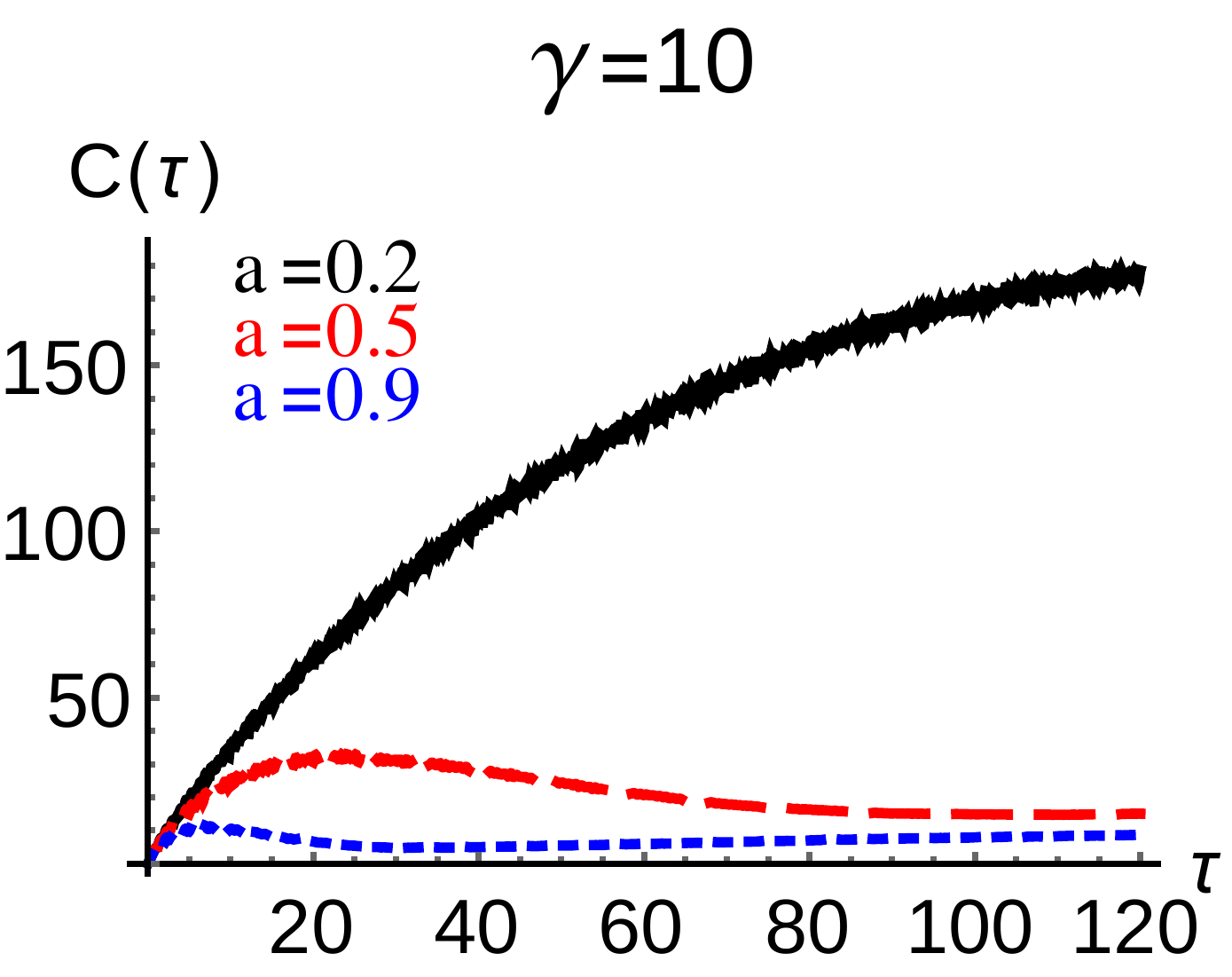}
\includegraphics[width=0.49\columnwidth]{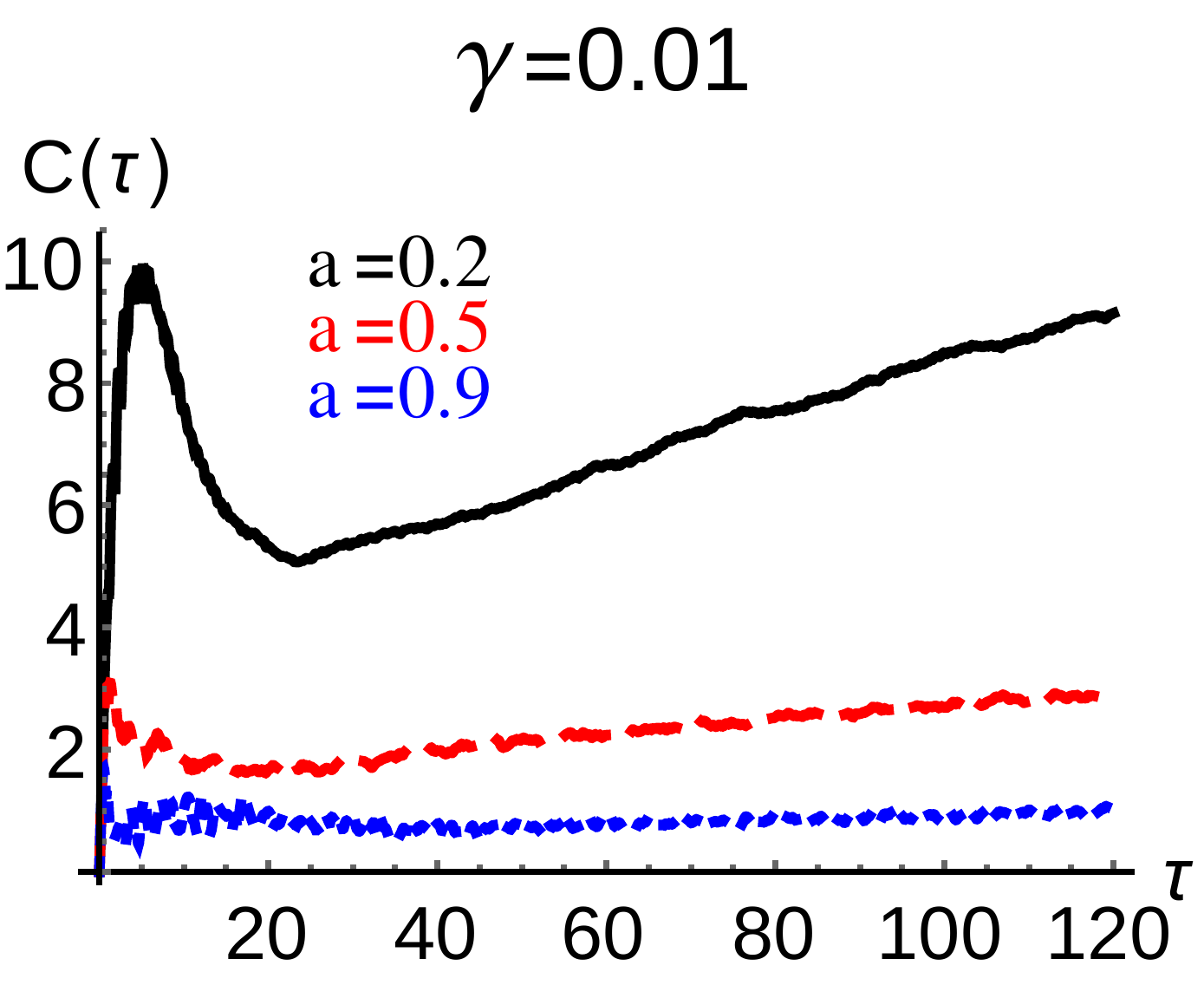}
\caption{The coherence $C$ of Eq. \eqref{coeren} as a function of the interaction time $\tau$ for
different values of the noise amplitude $a=$ $0.2$ (solid black line), $0.5$ (dashed red line)
and $0.9$ (dotted blue line), in the fast (left plot) and slow ( right plot) regime,
computed for a QW on a lattice with $N=500$ sites.}
\label{coeRTN}
\end{figure}
\par
What emerges from our analysis, so far, is that in the fast noise regime
[also referred to as the weak coupling regime since we are working with 
rescaled parameters, see Eq. (\ref{rescalp})] we see a transition from
quantum to classical diffusive behavior, while in the slow noise 
regime (strong coupling) we observe localization of the walker over 
few lattice sites. Since a classical
random walk has a Gaussian probability distribution over the line, we
may quantify the degree of classicality of the QW evolution by computing
the negentropy \eqref{negentropy}, i.e. by looking how far the QW
distribution is from a normal one. In Fig. \ref{negRTN} we report the
behavior of  $N_E$ as a function of time $\tau$ for different values of
the noise amplitude $a$ in the case of a QW subject to fast RTN (left
plot) and slow RTN (right plot).  The two regimes for
the autocorrelation function indeed identify two different behaviors for the
$N_E$: in the fast noise limit, the negentropy is smaller for long time
as the noise strength is increased, indicating that a transition toward
a classical, Gaussian probability distribution is induced by strong
noise. On the other hand, in the slow noise regime, the negentropy, after
an initial dynamics which is related to the noise amplitude, 
becomes almost constant, indicating that the probability distribution 
over the lattice changes only slightly as time is incresed.
\par
The effect of noise on the dynamics of the walker and the
appearance of a phase transition may be
analyzed in more details by looking at the time dependency of the variance $\sigma^2
(\tau)$ of the position of the walker, i.e. the spread of the particle 
over the lattice. The first plot in  Fig. \ref{sigmaRTN}. 
shows the 
variance for different values
of the noise amplitude in the case of $\gamma=10$. For small times, the
variance is quadratic in time, and it becomes linear at later times, a
signature that  a transition between quantum and classical diffusion has
happened. In fact, the curves may easily fitted by a quadratic function
for interaction times below a given threshold $\tau<\tau_c$ and by a 
linear one above this threshold. 
The transition time $\tau_c$ 
depends upon the value of the noise amplitude 
and it is larger for smaller value of $a$.
Notice that in our system we have decoherence without dissipation, 
such that the position variance provides a good indicator of the
transition. A different approach has been discussed for dissipative
systems \cite{garne}.
\par
The second plot shows the same quantity but for $\gamma=0.01$.
While for small noise amplitudes the walker can still propagate, as the
noise strength increases, the variance becomes linear, 
indicating that the walker is diffusing very slowly through
the lattice, thus confirming localization over few nodes around the initial
site. 
\par
The analysis of the dynamics of distribution over 
the lattice sites, as well as those of the negentropy and of 
the variance, only involves the diagonal elements of the 
density matrix. In order 
to gain more insight into the behavior of the system we study the 
time evolution of the full density matrix by analyzing
the dynamics of its coherence $C$, as defined in Eq.  \eqref{coeren}.
Fig. \ref{coeRTN} shows the dynamics of coherence for a noisy CTQW, in
the two regimes of fast (left plot) and slow (right plot) noise, for
different values of the noise amplitude $a$. In the fast regime, the
dependency on $a$ is clearly evident: the off diagonal elements of the
density matrix grows over time (here for a fixed interaction time
$\tau=120$) for small values of the noise amplitude. If we increase the
value of $a$, $C(\tau)$ starts decreasing after an initial growth, a
sign of decoherence induced by noise. 
In the slow noise regime (right panel), the coherence $C(\tau)$
initially increases and then 
{drop to a constant
for large values of $a$, while re-growth is seen for small 
noise amplitudes.} 
This indicates the survival of quantumness over time, as expected
for a system undergoing Anderson localization. The initial increase of
$C(\tau)$ is larger for smaller values of the noise amplitude $a$. We
also point out that the magnitude of $C(\tau)$ for slow noise is always
below the corresponding values for fast noise, in agreement with the
fact that the initially localized state tends to spread very little over
the lattice nodes.
\par
So far, we have shown that for a CTQW propagating in a disordered lattice
subject to RTN, two main typical long-time behaviors arises: the walker
can spread very slowly, staying localized over a  small fraction of the
total number of sites, or it can propagate through the graph with a
standard deviation proportional to the square root of time. One may
wonder whether this features depends upon the choice of the localized
initial condition, or they are more general characteristic of noisy QWs.
\begin{widetext}$ $ 
\begin{figure}[h]
\includegraphics[width=0.3\textwidth]{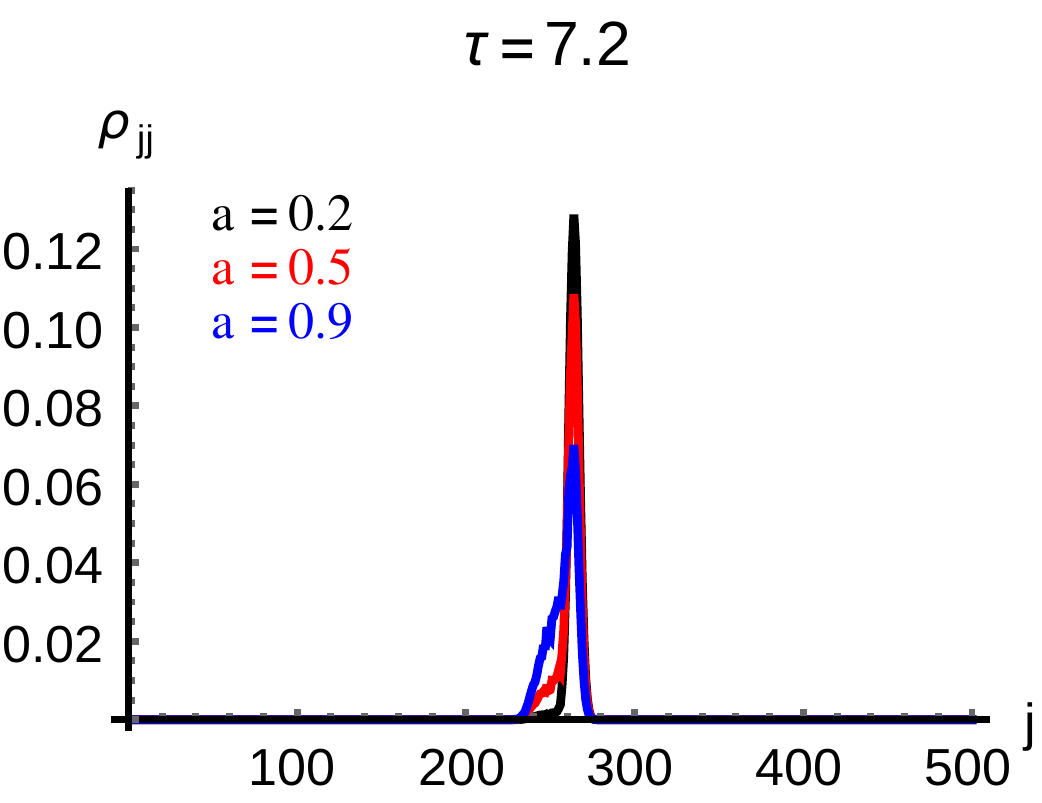}
\includegraphics[width=0.3\textwidth]{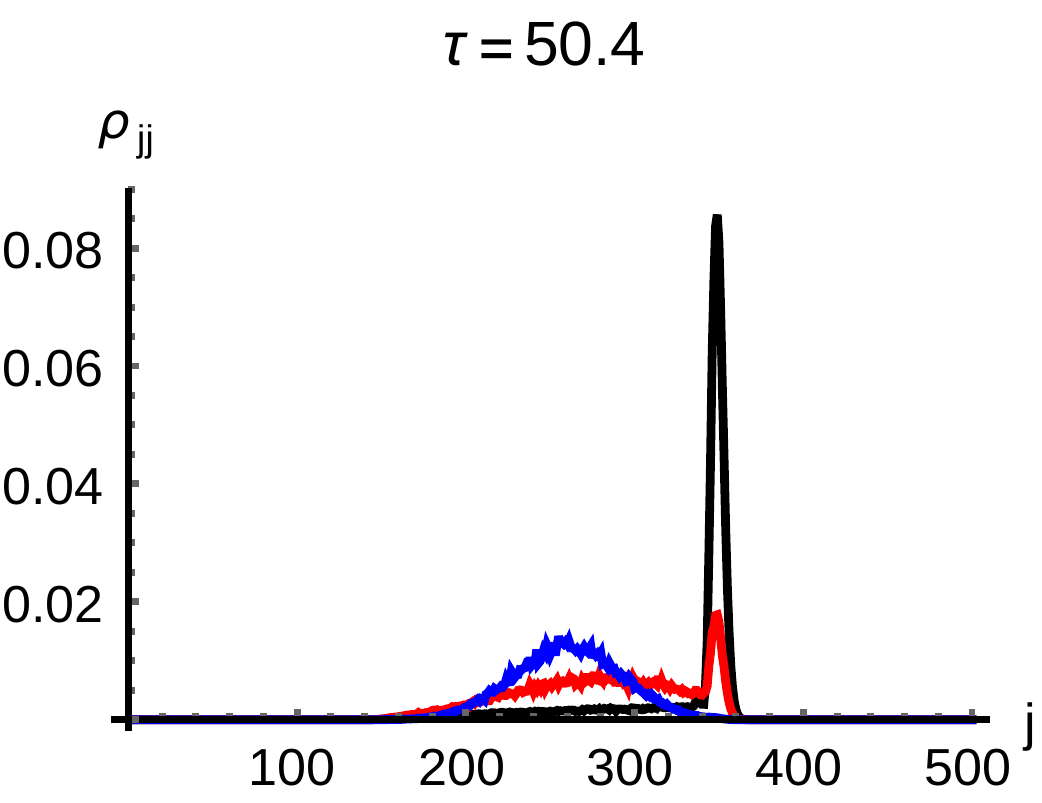}
\includegraphics[width=0.3\textwidth]{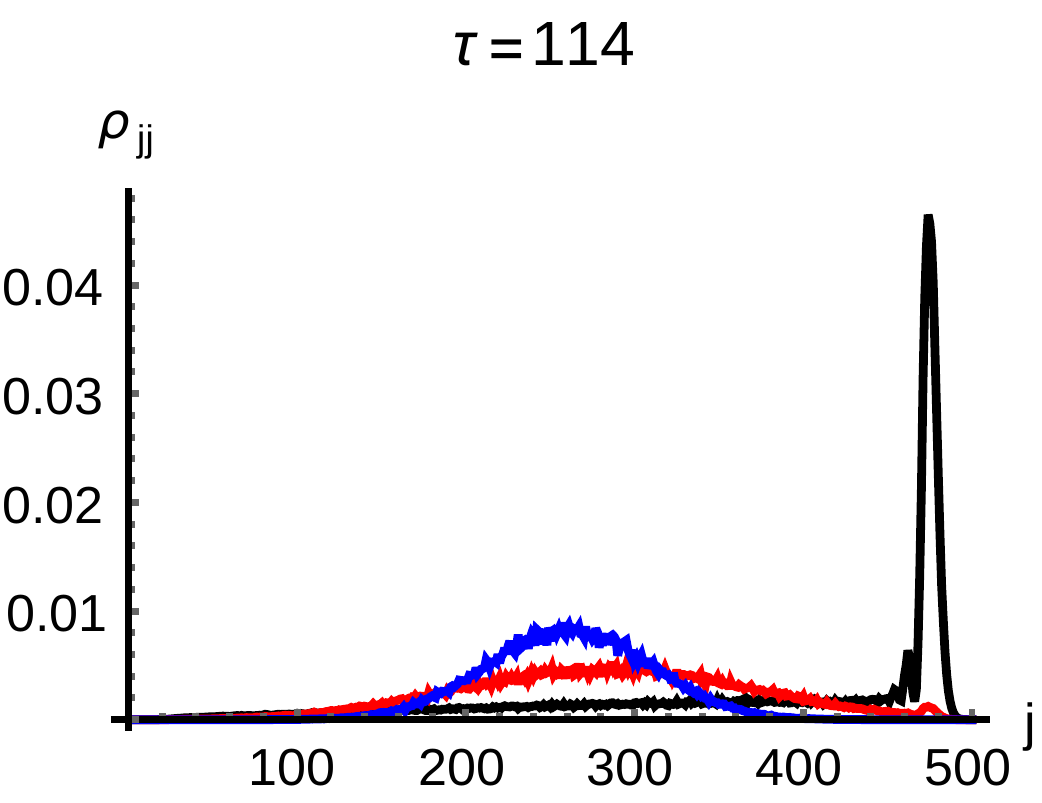}
\includegraphics[width=0.3\textwidth]{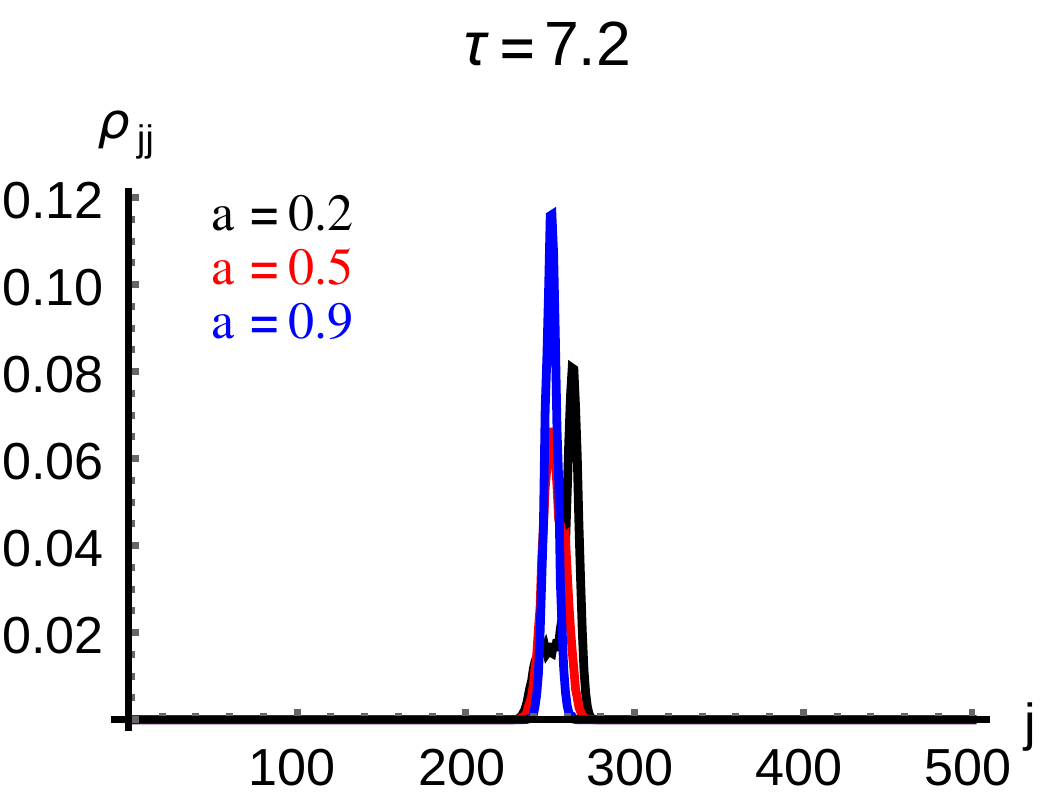}
\includegraphics[width=0.3\textwidth]{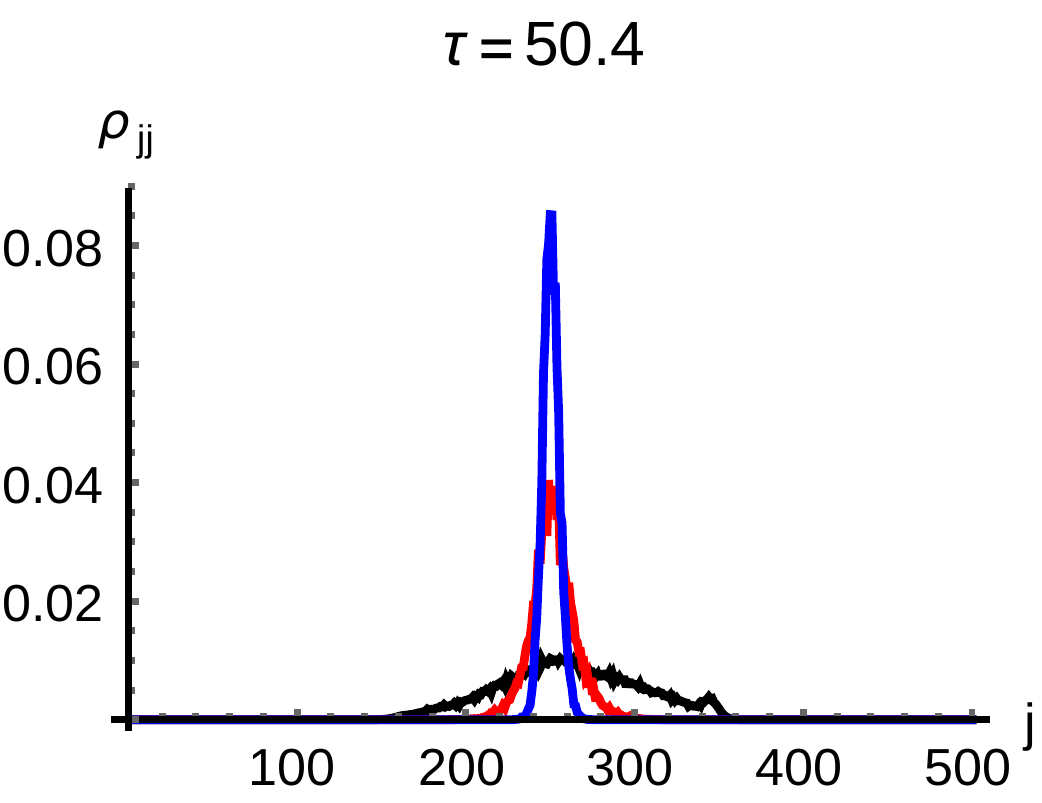}
\includegraphics[width=0.3\textwidth]{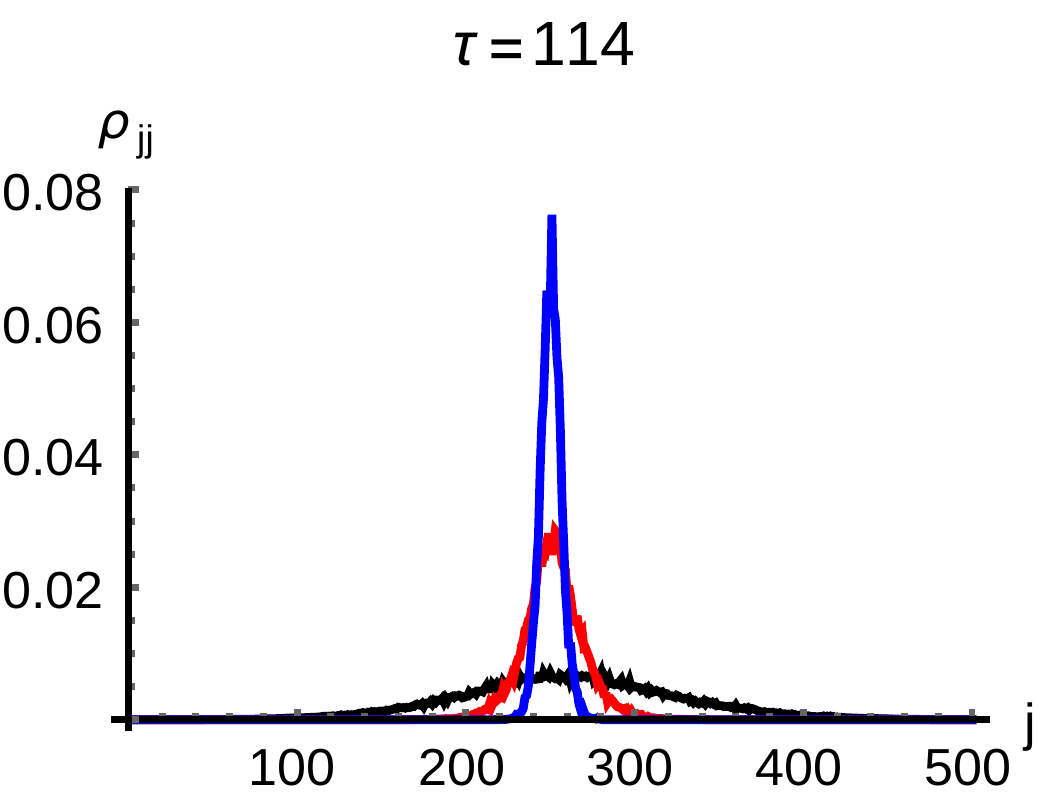}
\caption{Probability distribution of the particle over the lattice
nodes, for different values of the noise amplitude $a$ $=0.2$ (black line), $0.5$ (
red (light gray) line) and $0.9$ (blue (dark grey) line) and  interaction time
$\tau$, for a lattice of $N=500$ sites, subject to RTN with $\gamma=10$
(upper row) and $\gamma=0.01$ (lower row). The particle is initially
described as a Gaussian wavepacket with $\Delta=3$ and initial velocity
$k_0=\frac{3\pi}{2}$.} \label{gwpe}
\end{figure}
\end{widetext}
\subsection{An initial Gaussian wavepacket}
In order to better analyze the effect of dynamical disorder on the
quantum walk, we consider a different initial condition: instead of a 
localized initial state, we study the case of an initial Gaussian
wavepacket with a non-zero dimensionless velocity $k_0$ (in unit of
$\hbar$) and dimensionless standard deviation $\Delta$, such that the
initial (pure) state $\ket{\psi_G}$ may be written as:
\begin{equation}
 \ket{\psi_G}=\sum_{j=1}^N \left(\frac{1}{\sqrt{2\pi
 \Delta^2}}e^{-\frac{\left(j-\frac{N}{2}\right)^2}{2\Delta^2}}\right)
 ^{1/2}e^{-i k_0 j}\ket{j}.
 \label{gwpp}
\end{equation}
This initial condition allows us to imprint an initial momentum
distribution to the particle and investigate under which conditions 
transport phenomena over the graph is possible in the presence of noise. 
In Fig. \ref{gwpe} we report the probability distribution over the
lattice sites for different times, for different values of the noise
amplitude and the noise parameters, in analogy with Fig.
\ref{rtnprof1}. As before, we compare the dynamical behavior of the
walker subject to fast and slow RTN.  For small values of the parameter
$a$, the wavepacket moves away from its initial position during time,
indeed indicating transport through the graph.  The same features 
seen with a localized initial condition are found, thus indicating 
that the main features of the dynamics 
are imputable to decoherence and thus independent on the choice of the
initial condition.  Fast noise leads, indeed, to a Gaussian-like
probability distribution as the noise amplitude becomes larger, while
slow noise keeps the distribution localized.
\par
Fig. \ref{gwp2} shows the dynamical behaviors of the mean position
$\langle x(\tau)\rangle$ of the particle, its variance $\sigma^2(\tau)$
and the coherence $C(\tau)$, in the two noise regimes.  The main
difference here with respect the localized case is that the mean position
$\langle x\rangle$ changes with time for fast noise, moving away from
the initial position, indicating that not only diffusion is present, but
also drift.  From these results, we conclude that transport 
is possible if the strength of the noise $a$ is small,
otherwise the diffusive (or localized) behavior prevails, threatening the
possibility of transport.  This can also be confirmed by the analysis of
the variance in the two regimes.  As the noise strength is increased in
the weak coupling regime, the effect of decoherence manifests through
the spread of the wavepacket over the nodes, i.e. $\sigma^2(\tau)$
increases rapidly in time. In the other regime, instead, we see again
that as the value of $a$ is increased the wavefunction is localized with
a slowly varying variance. 
\par
The coherence of a quantum state initially prepared in a
superposition  decays faster for larger values of $a$ in both
regimes. However, for the explored values of the interaction time, 
superposition of states were preserved.
\par
Our results indicate that the timescale of the noise, i.e. its 
autocorrelation function, determines the qualitative behavior 
of the walker dynamics over the lattice. In the next section 
we are going to analyze and discuss how the two working regimes are 
related to the Markovian or non-Markovian nature of the quantum 
map describing the evolution of the walker.
 \begin{figure}[ht]
\includegraphics[width=0.48\columnwidth]{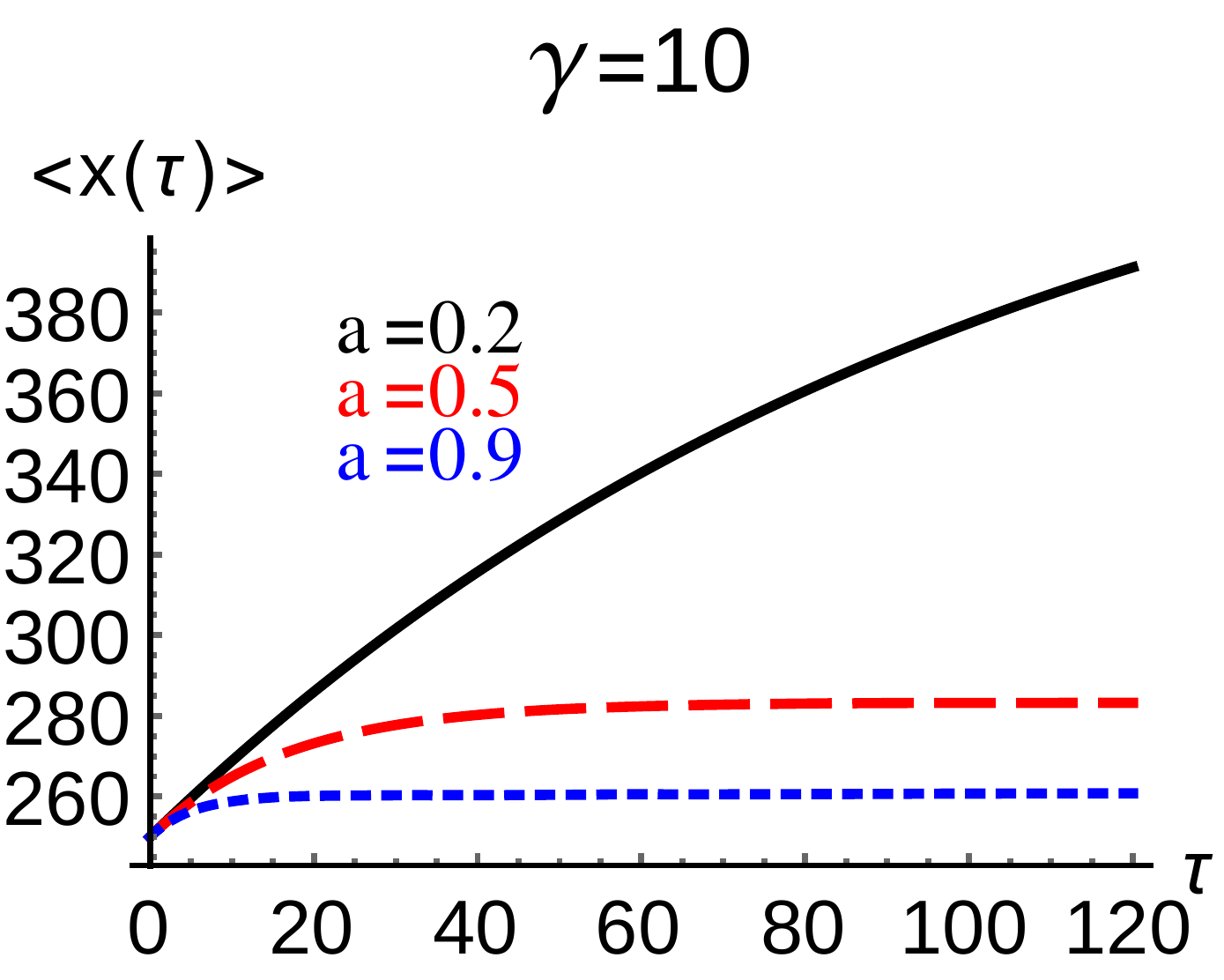}
\includegraphics[width=0.48\columnwidth]{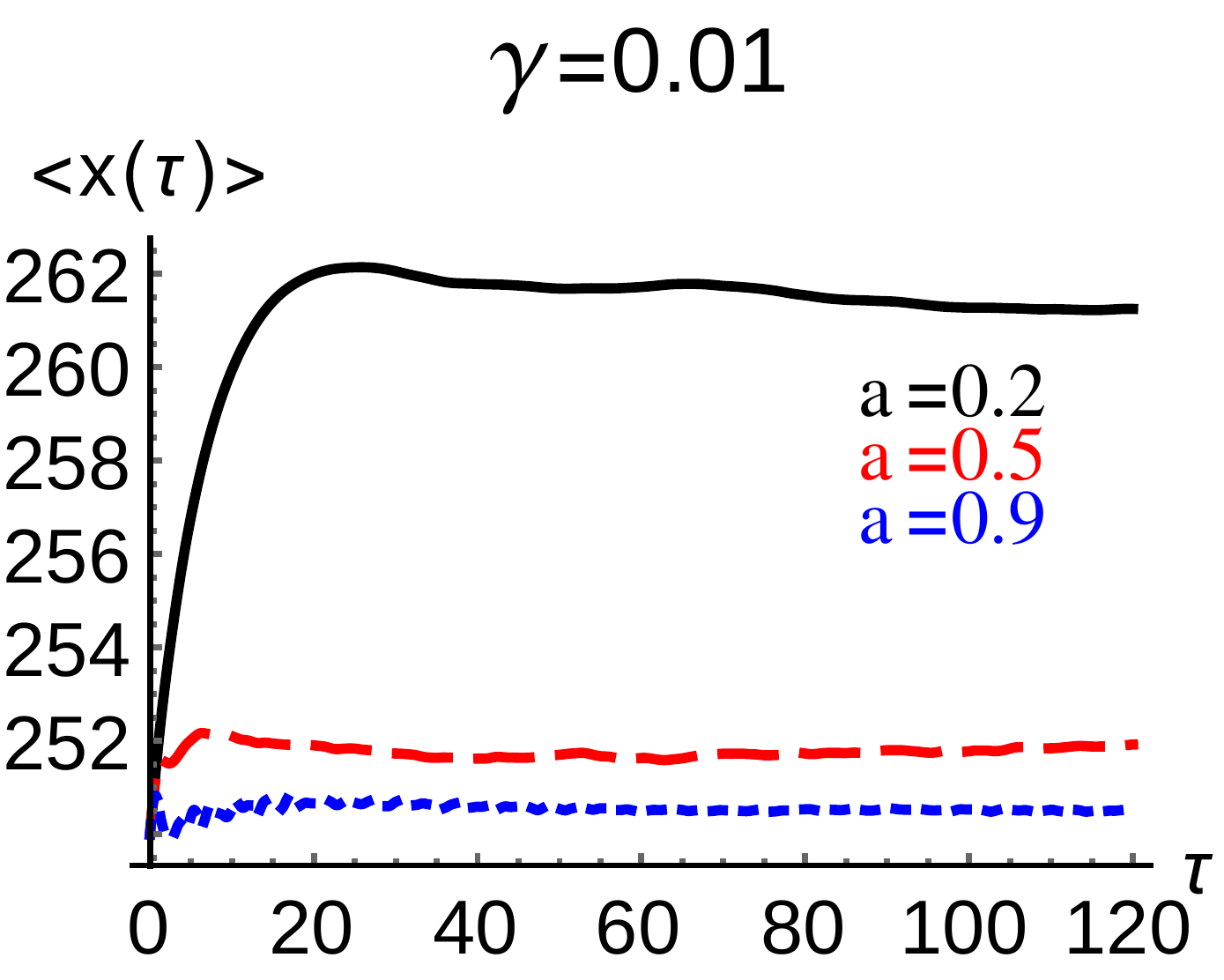}
\includegraphics[width=0.48\columnwidth]{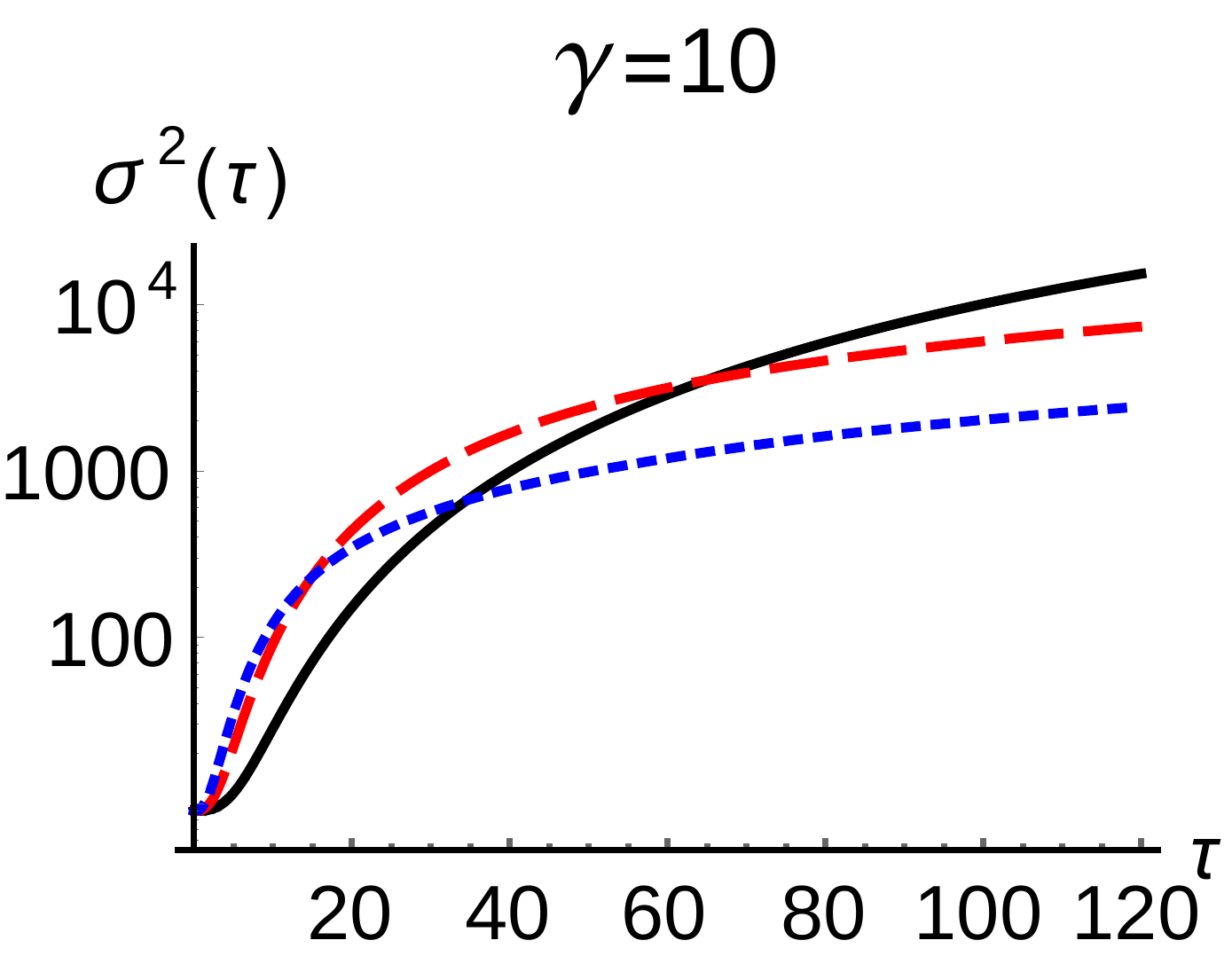}
\includegraphics[width=0.48\columnwidth]{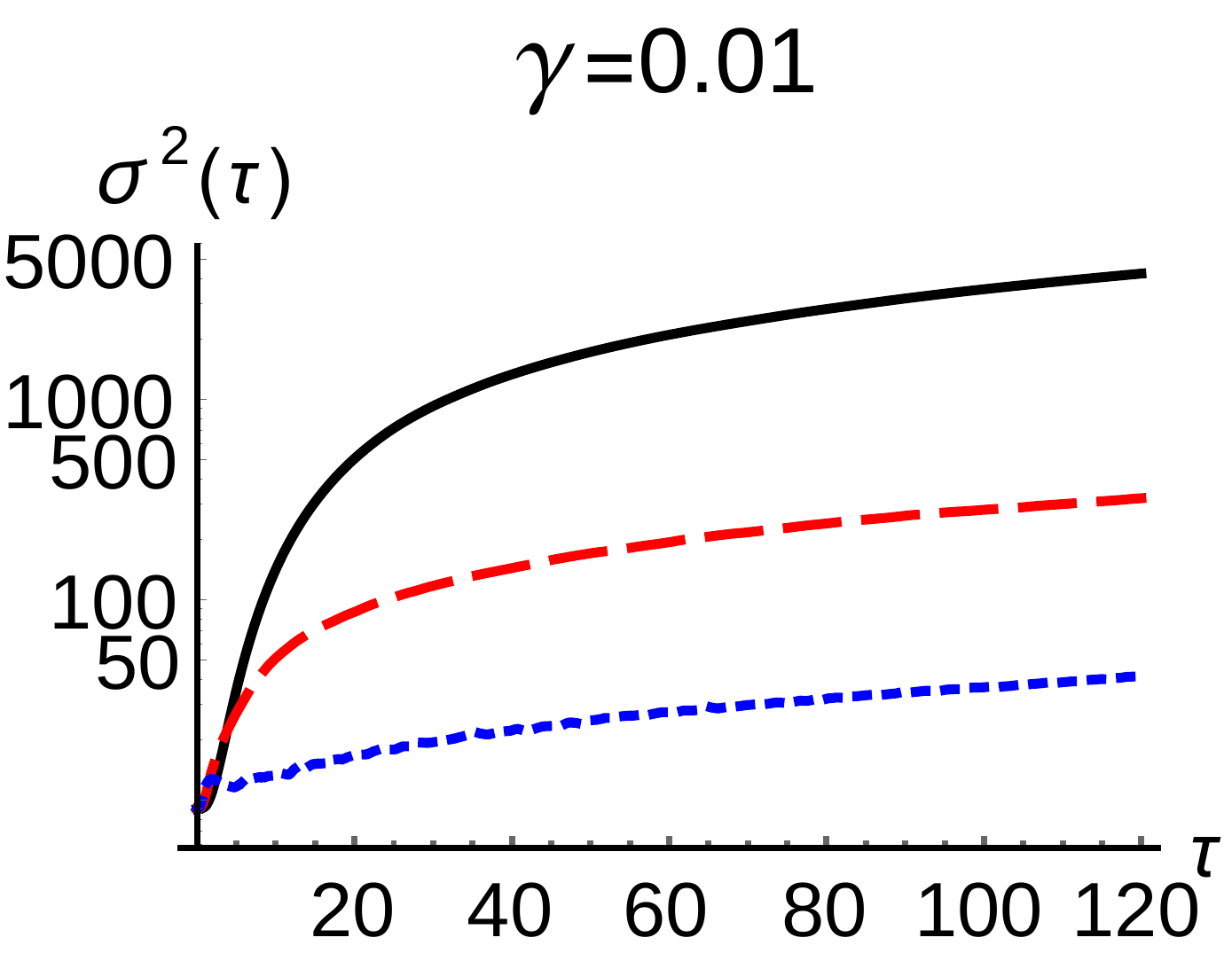}
\includegraphics[width=0.48\columnwidth]{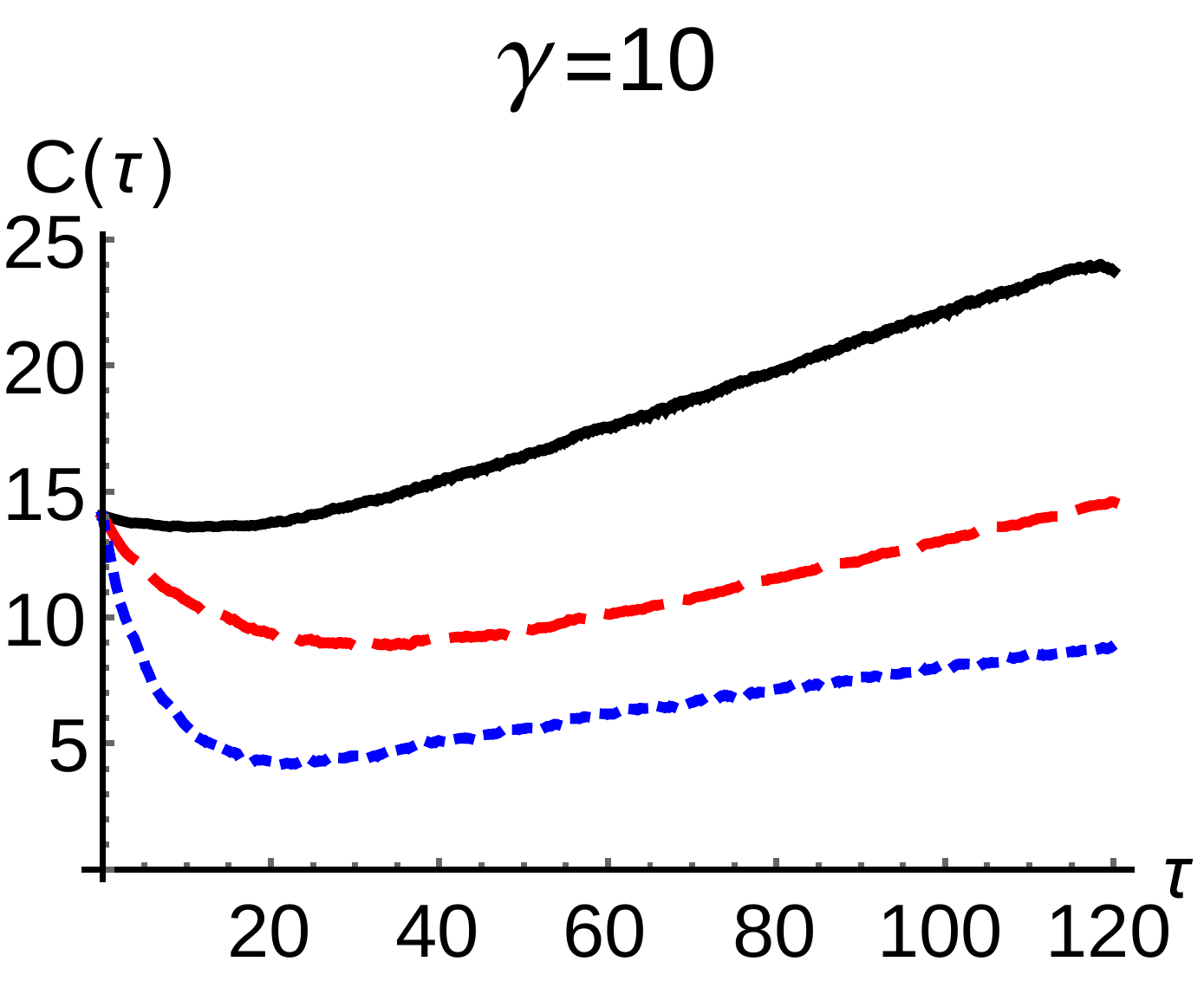}
\includegraphics[width=0.48\columnwidth]{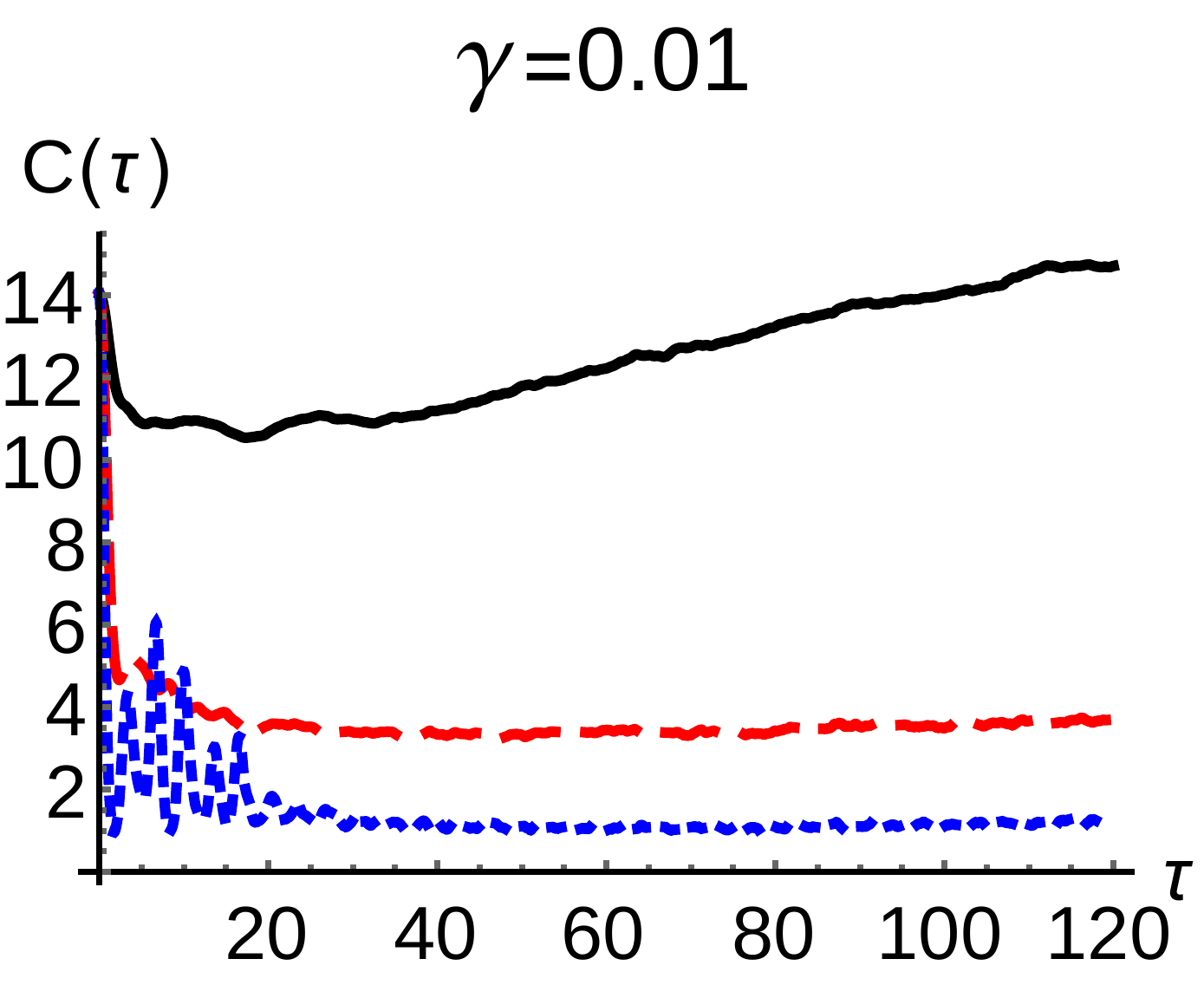}
\caption{From top to bottom: Mean value of the particle 
position $\langle x(\tau)\rangle$, variance $\sigma^2(\tau)$ and
coherence $C(\tau)$ as functions of the interaction time $\tau$, for different values of the
noise amplitude $a=$ $0.2$ (solid black line), $0.5$ (dashed red line)
and $0.9$ (dotted blue line), for a ring lattice of 500 sites, subject to RTN
with $\gamma=10$ (left column) and $\gamma=0.01$ (right column). The particle
is initially described as a Gaussian wavepacket with $\Delta=3$ and
initial velocity $k_0=\frac{3\pi}{2}$. } \label{gwp2}
\end{figure}
\section{Non-Markovianity of CTQWs}\label{sec:nm}
Non-Markovian quantum dynamics arises when memory effects become 
relevant and the future evolution of a quantum system does not 
only depend upon its present, but it is instead determined by 
its full or partial past history.
\par
A slowly decaying autocorrelation function for the environmental noise
may be intuitively associated with memory effects in the environment,
while fast decaying, delta-like, autocorrelation, is 
usually associated to Markovian dynamics.  In order to check 
whether this connection is true, we assess the non-Markovian
character of the dynamical map by looking at violation 
of Eq. \eqref{mnm} by some given intial states, and also by
studying the evolution of the trace distance between suitably chosen 
pairs of states. Both techniques have limits, i.e. may not provide 
full characterization of the dynamical map, since for CTQWs 
one cannot span all intermediate times $\tau_1$ to check validity of Eq. \eqref{mnm} or 
span the full Hilbert space in looking for states that experience 
information backflow. Yet, we may obtain numerical evidence (violation of 
equality \eqref{mnm}) for the non-Markovianity of the evolution in 
the presence of slow noise. 
\begin{figure}[th]
\includegraphics[width=0.49\columnwidth]{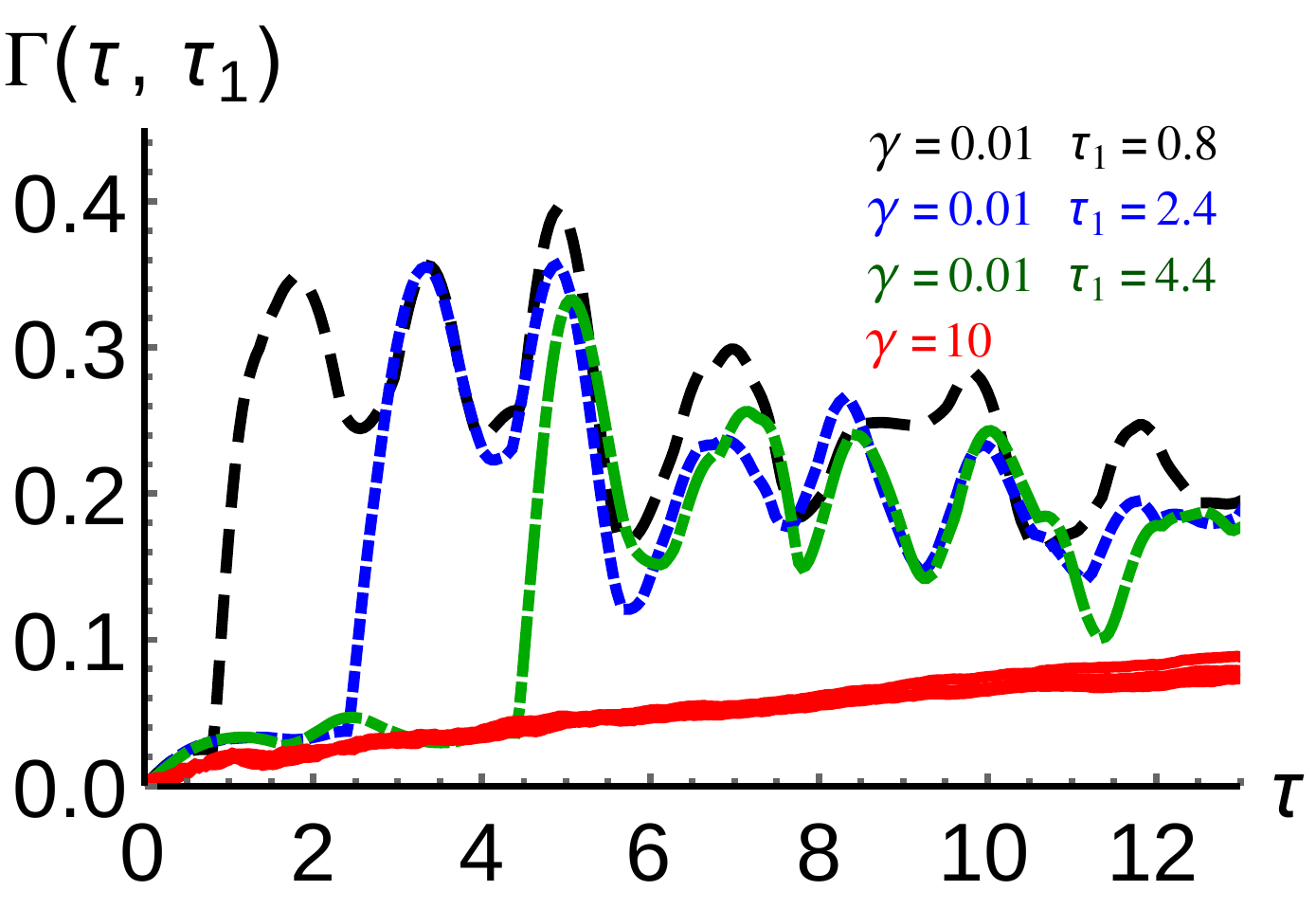}
\includegraphics[width=0.49\columnwidth]{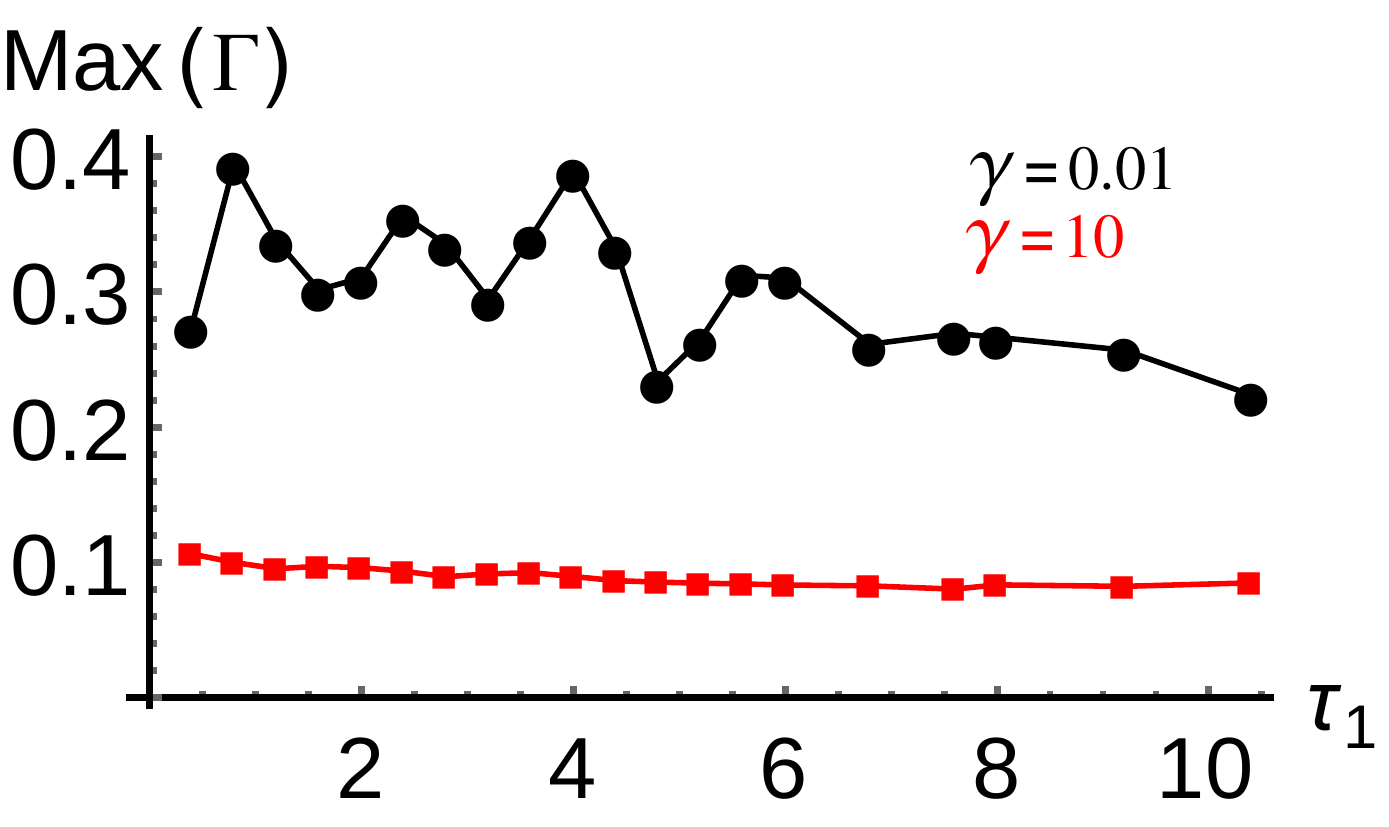}
\caption{Left: Trace distance $\Gamma(\tau,\tau_1)$ 
as a function of the interaction time $\tau$ in
the case of slow (dashed black, dotted blue and dot-dashed green lines) and fast (solid red lines) noise. The different
curves are for different values of intermediate time $\tau_1$ in Eq.
\eqref{gam}. The particle is initially in a localized state.
Right: Maximum of $\Gamma(\tau,\tau_1$) over time $\tau$ as a function of $\tau_1$. The black points are 
for $\gamma=0.01$ and the red squares for $\gamma=10$.}
\label{nmfig2}
\end{figure}
\begin{figure}[ht]
\centering
\includegraphics[width=0.49\columnwidth]{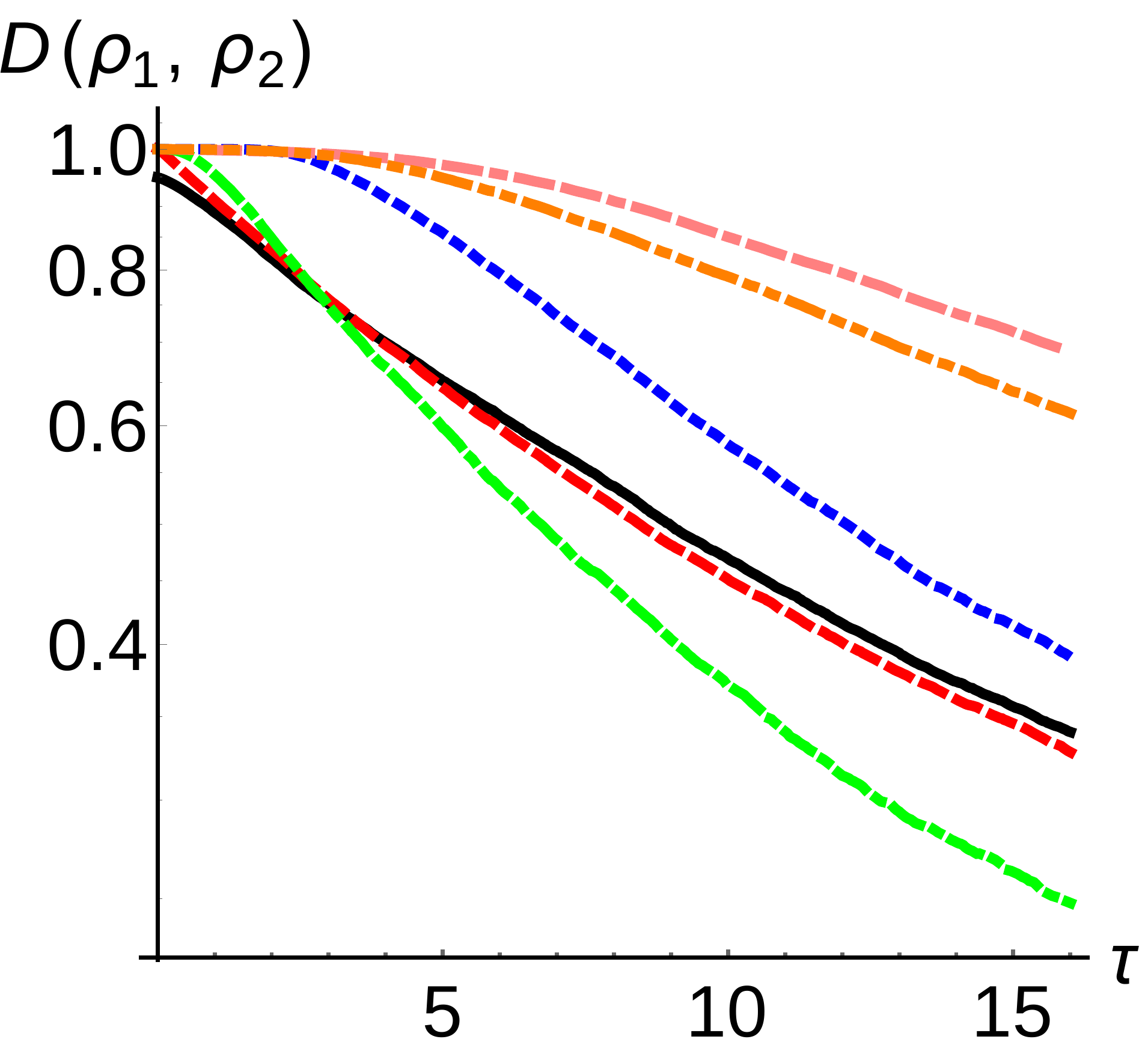}
\includegraphics[width=0.49\columnwidth]{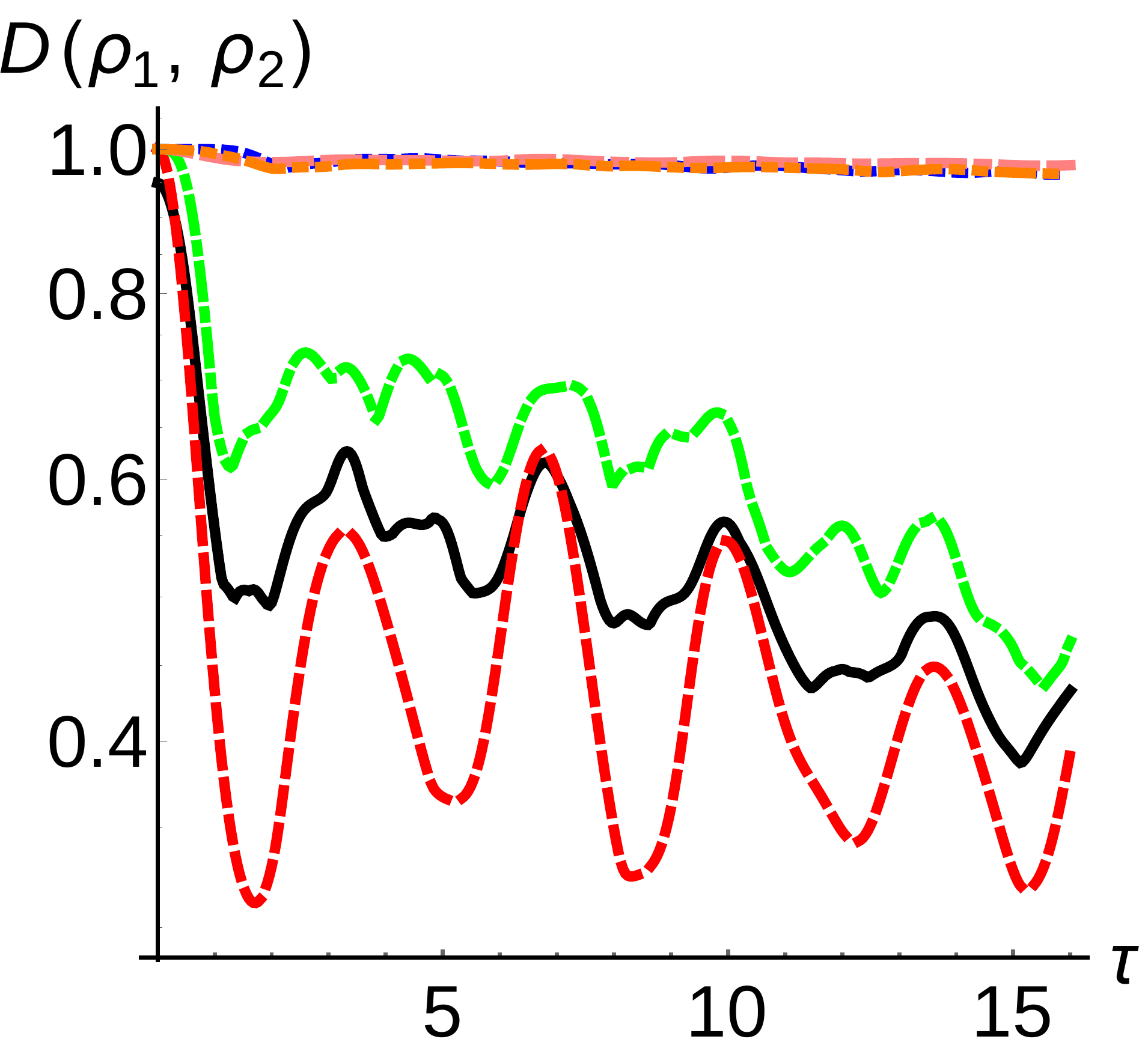}
\caption{
Analysis of the trace distance for selected initial pairs of states.
Trace distance for particle subject to RTN, for fast (left) and slow
(right) noise. Different colors refer to different choices of the
initial pairs of pure states: 
(solid black line) $\ket{x_0}, \ket{\psi_G(x_0, k_0)}$, (dashed red line)
$\ket{\psi_G(x_0,k_0}, \ket{\psi_G(x_0, k_0+ \frac{20\pi}{N})}$, (dot-dashed green line)
$\ket{x_0}, \frac{\ket{x_0+3}+\ket{x_0-3}}{\sqrt{2}}$, (dotted blue line)
$\ket{x_0}, \ket{x_0+10}$, (dot-dash-dash pink line) $\ket{\psi_G(x_0,k_0}, \ket{\psi_G(
x_0+20, k_0+\frac{20\pi}{N}))}$, (dot-dot-dashed orange line) $\ket{x_0},\ket{\psi_G( x_0+6,
k_0)}$, where $x_0=N/2$, and $\ket{\psi_G(x_0,k_0}$ is the initial
Gaussian wavepacket  in Eq. \eqref{gwpp} with initial position $x_0$,
velocity $k_0=3\pi/2$ and standard deviation $\Delta=4$.}
\label{nmfig3}
\end{figure}
\par
In order to prove the violation of Eq. \eqref{mnm}, 
we consider a suitable initial state $\rho_0$ and evaluate
the trace distance \eqref{tdis} between the state  
obtained  by applying the full map $\rho (\tau)= T(\tau,0)\rho_0$ 
and the state  resulting from the composition
$\rho^\prime (\tau) =T(\tau,\tau_1)T(\tau_1,0)\rho_0$, i.e.
\begin{align}
\Gamma(\tau,\tau_1)=
D\Big(T(\tau,0)\rho_0,\, T(\tau,\tau_1)T(\tau_1,0)\rho_0\Big).
\label{gam}
\end{align}
In the left panel of 
Fig. \ref{nmfig2} we show the behavior of $\Gamma(\tau,\tau_1)$ 
as a function of time in the fast 
and slow 
noise regime, starting from a localized initial condition. For a 
given value of $\gamma$, the different lines correspond to different 
values of the intermediate time $\tau_1$ in Eq. \eqref{gam}. As it is 
apparent from the plot, in the fast noise regime $\Gamma(\tau,\tau_1)$ 
is close to zero at any time and for any choice of the intermediate
$\tau_1$, i.e. no differences appear between the full and composed 
dynamical maps. The fact that the trace distance is not strictly 
zero is imputable to the accumulation of numerical noise, since
we are averaging over a finite number of realizations of the 
stochastic processes $\{g_j(t)\}$ and not performing the true ensemble average
as in Eq. \eqref{evolution}.
\par
On the contrary, strong differences in the dynamics are revealed 
when we consider the slow noise regime [i.e. strong coupling, see Eq.
(\ref{rescalp})], indicating that the dynamical map involves 
memory effects for slowly fluctuating environments 
with long-lasting correlations.
This is confirmed by the results reported in the right panel of 
Fig. \ref{nmfig2}, which shows the  maximum of the trace distance 
$\Gamma$ over time, for different values of $\tau_1$, in both the 
fast and slow noise regime. When the dynamical evolution is split
at a certain value of $\tau_1$ in the slow noise regime, the 
difference between the full and composed dynamical maps are apparent, 
while in the fast regime ($\gamma=10$), values of the maximum are 
compatible with the numerical noise.
\par
Notice that this is a property of the map, and thus it 
may not reveal
itself for all initial states. Indeed, in our case, we have detected
clear violation of Eq. (\ref{mnm}) using localized initial conditions, 
while starting from an initial Gaussian wavepacket (with or without a 
velocity distribution) lead to small value $\Gamma(\tau,\tau_1)$, i.e. 
to nearly divisible evolution.
\par
On the other hand, the non-Markovian character of the dynamical map
induced by slow noise is confirmed by analyzing the behavior of the
trace distance between initial pairs of states, as required by BLP
measure.  Oscillations in time of  $D(\rho_1(\tau),\rho_2(\tau))$  
for some given initial pair $\rho_1(0)$, $\rho_2(0)$ provide evidence for 
information backflow to the quantum system from the surrounding
environment. It is thus necessary and sufficient to find one initial
pair for which the trace distance is non-monotonic to prove that the
dynamical map is non-Markovian, even if we cannot make a qualitative
statement about the degree of non-Markovianity.  
Indeed, in the case of slow RTN, it is quite simple to sample the state
space and find an initial pair of states for which the trace distance 
shows revivals during time evolution, as shown in Fig. \ref{nmfig3}. 
On the other hand, 
we could not find any initial pair leading to non-monotonic behavior 
in the fast noise regime, a fact suggesting that the map 
may be Markovian, even though this does not prove it,  
since we are not able to check all possible initial states. Still 
this result is in agreement with the analysis of 
the composition equality in \eqref{mnm}, thus indicating the lack 
of memory effects in the fast noise regime and, in turn, 
provides strong indication in that direction.
Overall, our results shows that in the slow noise (strong coupling) 
regime, the dynamics is non-Markovian: memory effects are important and
allows one to observe 
information backflow. At the same time, there are robust numerical
evidences that the fast noise (weak coupling) regime corresponds to a 
Markovian dynamics. These results are also in agreement with our previous 
results for simpler systems, e.g. concerning the non-Markovianity 
of quantum maps describing the interaction of qubits with
classical fluctuating fields \cite{benedenm}.
\begin{table}[h!]
\begin{tabular}{l|c|c|}
Value of $\gamma$ &$\gamma\ll 1$ & $\gamma\gg 1$\\
 \hline
Regime& Slow noise  & Fast noise\\
 \hline
(coupling) &(strong coupling)&(weak coupling)\\
 \hline
 Dynamics& Localized & Transition to classical\\
 &  & diffusion\\
 \hline
Memory, see Eq. \eqref{mnm}& Yes& No\\
\hline
BLP measure & Non-Markovian & Markovian\\
\hline
\end{tabular}
\caption{{Summary of the main features of two dynamical regimes.}}
\label{t:1}
\end{table}
\section{Conclusions}\label{sec:concl}
In the last decade there have been several proposals to implement
quantum walks in different systems and different kind of lattices
\cite{wang14}, also addressing scalability and feasibility in realistic
conditions.  In this framework, it becomes crucial to have more
realistic theoretical models, which take into account the effects of
noise and assess the residual quantumness of the systems. In fact,
imperfections in the fabrication of the lattice may introduce randomness
in the tunneling energy of the walker, thus inducing detrimental
fluctuations that may, or may not, destroy the quantum effects in the
system.  
\par
In this paper we have studied in details the dynamics of noisy
one-dimensional CTQWs.  Defects and disorder in the lattice has been
described as stochastic classical processes governing the off-diagonal
elements of the Hamiltonian, in order to describe
fluctuations in the tunneling amplitudes between neighboring sites. The
walker dynamics has been then computed as an ensemble average over
possible realizations of the noise.  We found that, depending on the
ratio between the autocorrelation time of the noise and the coupling
between the walker and the external environment generating the noise,
two different dynamical regimes appear.
\par
If the walker is strongly coupled to its environment, the
resulting slow noise confines the walker 
into few lattice nodes. On the contrary, a weakly coupled 
walker subject to fast noise is driven through a transition 
from quantum ballistic diffusion to a classical
diffusive propagation over the lattice. The peculiar features
of the two dynamical regimes have been confirmed by analyzing
the variance of the particle position, the negentropy of 
the distribution, and the overall coherence of the full density matrix. 
We have analyzed different initial conditions for the walker, 
either localized or a Gaussian wavepacket, thus also exploring 
the conditions under which we have information transfer through
the lattice. We found that transport is possible if the amplitude $a$ 
of the noise is small, otherwise the diffusive (or localized) 
behavior prevails.
\par
Upon analyzing the properties of the dynamical map, we have established
a connection between the behavior of the walker in the slow
noise/strong coupling regime and the non-Markovian character of the
evolution. In particular, we have shown that in this regime the dynamics
cannot be written as the composition of two memoryless universal dynamical maps
, i.e. violates Eq. \eqref{mnm}, 
a signature that memory effects are important, and
that the information lost because of noise may {\em flow back} to the
system.  In the fast noise/weak coupling regime, numerical evidences
strongly suggest the Markovianity of the quantum map, even if a
conclusive proof is not currently available.
\par
By tuning the ratio between the memory parameter of the noise 
and the coupling with the walker, it is possible to move continuously 
from one dynamical behavior to the other and observe the corresponding
transition between the two dynamical phases, see Table \ref{t:1}. 
This is a relevant 
feature, since the chance of controlling the transition 
between different evolutions 
would serve as guidelines for reservoir engineering, where 
noise may be exploited to enhance some desired dynamical features.
\section*{Acknowledgments}
CB and MGAP thank Dario Tamascelli and Kimmo Luoma for discussions and useful 
suggestions. This work has been supported by EU through the 
Collaborative Projects QuProCS (Grant Agreement 641277), by 
UniMI through the H2020 Transition Grant 14-6-3008000-625 and by UNIMORE through FAR2014.

\end{document}